\documentclass[aps,prf,preprint,groupedaddress]{revtex4-2}

\usepackage{graphbox}
\usepackage{xcolor}
\newcommand{\qq}{\underline{q}}
\newcommand{\qo}{\underline{h}}
\newcommand{\qt}{\underline{q}}
\newcommand{\xx}{\underline{x}}
\newcommand{\uu}{\underline{u}}
\newcommand{\UU}{\underline{U}}
\newcommand{\tM}{\tilde{M}}
\newcommand{\pphi}{\underline{\phi}}

\newcommand{\commentYF}[1]{\textcolor{blue}{YF:}\textcolor{blue}{#1}}
\newcommand{\commentGB}[1]{\textcolor{red}{GB:}\textcolor{red}{#1}}

\usepackage{amsmath}
\usepackage{geometry}
\usepackage{fancyhdr}
\usepackage{natbib}
\usepackage{multirow}
\usepackage{comment}

\usepackage{graphbox}
\usepackage{xcolor}

\begin{document}

\title{On the relationship between the base pressure and the velocity in the near-wake of an Ahmed body}

\author{B\'ereng\`ere Podvin}
\affiliation{EM2C, CentraleSupélec, CNRS, Universit\'e Paris-Saclay}
\author{St\'ephanie Pellerin}
\affiliation{LISN, CNRS, Universit\'e Paris-Saclay}
\author{Yann Fraigneau}
\affiliation{LISN, CNRS, Universit\'e Paris-Saclay}
\author{Guillaume Bonnavion }
\affiliation{École Nationale Supérieure de Mécanique et d'Aérotechnique, Poitiers}
\author{Olivier Cadot}
\affiliation{School of Engineering, University of Liverpool}

\begin{abstract}
We investigate the near-wake flow of an Ahmed body  which
is characterized by switches between two asymmetric states that are mirrors of each other
in the spanwise direction. The work focuses on the relationship between the base pressure distribution and the near-wake velocity field. 
Using direct numerical simulation obtained at a Reynolds number 
of 10000 based on incoming velocity and body height as well as Bonnavion and
Cadot's experiment \cite{kn:bonnavion18}, we perform Proper Orthogonal Decomposition (POD) of the base pressure field.
The signature of the switches is given by the amplitude of the most energetic, antisymmetric POD mode. However switches are  also characterized by a global
base suction decrease, as well as deformations in both vertical and lateral directions, which all correspond to large-scale symmetric modes.
Most of the base suction reduction is due to 
the two most energetic symmetric modes. 
Using the linear stochastic estimation technique of \cite{kn:expif18},
 we show that the large scales of the near-wake velocity field  can be
recovered to some extent from the base pressure modes.
Conversely, it is found that the dominant pressure
modes and the base suction fluctuation can be well estimated from the POD velocity modes of the near-wake.
\end{abstract}

\maketitle

\section{Introduction}

Bluff bodies are characterized by a large pressure drag, the reduction of which represents
a significant industrial challenge in order to reduce vehicle emissions or to increase the range of electric vehicles.
The present study focuses on the flat-backed Ahmed body, which represents a simplified model to study 
the aerodynamics of trucks, SUVs and other flat-backed vehicles.
For sufficiently large ground heights, the near-wake of the body is characterized by
bistability, corresponding to asymmetric (reflection-symmetry-breaking or RSB) mirror states \cite{kn:grandemange13}.
The variations of the pressure drag are correlated with the occurrence of intermittent switches between 
these two states, during which the flow becomes briefly symmetric.
It has been shown that reducing the deviation of the wake leads to a drag reduction, which has motivated
several control attempts (\cite{kn:barros14}, \cite{kn:li16}, \cite{kn:evrard16}, \cite{kn:brackston16}, \cite{kn:plumejeau19}).

A central question is to determine how the different velocity motions contribute to the total drag. 
A large number of studies have therefore focused on the description of the flow dynamics.
Beside the large-scale switches between the quasi-stationary states, which take place on a time scale
of O(1000) convective time units based on the incoming flow velocity
and body height, the flow is also characterized by three-dimensional vortex shedding, with a characteristic
frequency of 0.2, as well as 
deformations of the recirculation zone associated with low frequencies in the range 0.05-0.1 in convective time units. 
These time scales have been identified in experiments \cite{kn:pavia18}, \cite{kn:perry16}  and confirmed by 
numerical simulations \cite{kn:lucas17}, \cite{kn:evstafyeva17},  \cite{kn:rao18}, \cite{kn:dallalonga19}, \cite{kn:fan20}. We note that all times will be expressed in convective time units throughout the paper.

Understanding the connection between the base pressure and the near-wake flow is a key ingredient of
a successful control strategy. Stochastic estimation, originally developed in a conditional average framework, 
 can provide useful insight into this relationship. 
Linear Stochastic Estimation (LSE) was first  developed by Adrian \cite{kn:mo&adrian} 
to identify coherent structures in a turbulent boundary layer, then its 
interest for control purposes was quickly made apparent \cite{kn:taylorglauser02}, which spurred on a number of variants.

Several implementations of LSE are based on the combination of linear stochastic estimation with Proper Orthogonal Decomposition (POD).
Bonnet et al. \cite{kn:bonnet94} first proposed an estimate 
based on a low-dimensional representation of the flow.  
We note that a connection can be made between linear stochastic estimation and the extended Proper Orthogonal Decomposition proposed by Bor\'ee \cite{kn:boree}.
Extensions of the method include spectral linear stochastic estimation, which has been applied to shear layers  \cite{kn:citriniti2000} as well as jets \cite{kn:tinney07}, 
and multi-time delay stochastic estimation, which has been applied 
to separated flows such as  a backward-facing step
(\cite{kn:hudy07}, a cavity shear layer \cite{kn:lasagna13},
or the wake of a bluff body \cite{kn:durgesh10}.  
In \cite{kn:expif18} a new LSE-POD variant was proposed in which POD is applied to both the conditional and unconditional variables.
The method was applied to a turbulent boundary layer and provided 
 a highly resolved velocity estimate in both space and time
from the combination of  
spatially resolved measurements with a low time resolution 
and and time resolved, but spatially sparse measurements.   

In many studies,  wall pressure measurements are used as the conditional variable to estimate the velocity field, 
since wall pressure sensors are non-intrusive and can easily be fitted on a body. 
However,  it can be argued that the 
inverse problem is also of interest, as it allows identification of the velocity patterns that create pressure fluctuations, following boundary layer 
studies \cite{kn:chang99}, \cite{kn:ferreira21}.

The purpose of the present paper is to investigate how the near-wake velocity field  depends on the base pressure,   
but also to determine how the fluctuations of the base pressure field are related to the dominant wake motions. 
To do this, we consider a direct numerical simulation
of the flow behind an Ahmed body at a moderate Reynolds number $Re=10^4$. We
 apply POD  to extract the near-wake velocity modes, as well as 
the dominant base pressure structures, which we  
compare with the experimental results
of Bonnavion and Cadot \cite{kn:bonnavion18}. 

The paper is organized as follows: after describing the numerical and experimental settings, we carry out
 POD analysis of the base pressure and of the near-wake velocity field. 
Using POD-based linear stochastic estimation,  we 
investigate the relationship between the most energetic POD pressure modes and the dominant POD velocity modes in the near-wake.

\section{Methodology}

\subsection{Numerical simulation}

The characteristics of the simulation have been given in  \cite{kn:podvin21}.
The dimensions of the squareback Ahmed body are  the same as in  Evrard et al. \cite{kn:evrard16} 
, $L = 1.124m, H =0.297m$, $W = 0.350m$. 
The ground clearance (distance from the body to the lower boundary of the domain) was taken equal 
to $C=0.334 H$ in the simulation.
The code SUNFLUIDh, which is a finite volume solver, is used to solve the incompressible Navier-Stokes equations.
The Reynolds number based on the fluid viscosity $\nu$,  incoming velocity $U_\infty$ 
and Ahmed body height $H$ is $10^4$.
The temporal discretization is based on
a second-order backward Euler scheme, with implicit treatment of the viscous terms and explicit representation of the convective terms (Adams-Bashforth scheme). The divergence-free velocity and pressure fields are obtained from a incremental projection method \cite {kn:goda79}.
We use (512 × 256 × 256) grid points in, respectively, the longitudinal
direction x, the spanwise direction y, and the vertical direction z, with a time step  set to $\Delta t= 5\ 10^{-4}$ (the CFL number never exceeded $0.4$).
The flow was integrated over more than 500 convective time units.

\subsection{Constitution of the POD Dataset}
The investigation described in this paper relies on POD analysis, which
is detailed  in the Appendix.
We apply the method of snapshots \cite{kn:siro87} to both the fluctuating pressure $c_p'$ and velocity fields $\uu'$ where $c_p(t)=C_p+c_p'(t)$, $\uu(t)=\UU+\uu'(t)$ and capital letters refer to as time averaged values. The pressure $p$ has been translated into a pressure coefficient defined as $c_p=2\frac{p-p_\infty}{\rho U_\infty^2}$ where $\rho$ is the air density, $p_\infty$ and  $U_\infty$ respectively the free stream pressure and velocity. The velocity $\uu$ is given in $U_\infty$ units. 
Since the quasi-stationary states break the spanwise symmetry,
statistical symmetry in the set of snapshots was enforced by adding
to each snapshot extracted from the simulation the image of that snapshot
through a reflection symmetry with respect to the mid-vertical plane. This is the same symmetrization procedure used in \cite{kn:podvin20}. 
Results shown in this paper were based on a set of 600 snapshots taken every convective unit,
300 of which are extracted from the simulation, and 300 are obtained through a reflection symmetry. 
The time span corresponds to that used for POD in \cite{kn:podvin21}, that is 300 convective time units, 200 of
which correspond to the duration of a switch. 
The last 200 time units were used to test the estimation procedure for both the pressure and the velocity fields.
 Comparison with another set of snapshots based on $2 \times 200$ samples
acquired spanning the duration of the second switch did not show any significant differences.
The normalized amplitudes of the $k-th$ modes for the pressure and velocity components will be respectively referred
to as $a_{k}^p$ and $a_k^v$. 
We will refer to the corresponding estimated amplitudes as $a_k^{pe}$ and $a_k^{ve}$. 

\section{Pressure field}

POD was first applied to the fluctuating pressure field on the base of the body (see Appendix) :
\begin{equation} 
c_p'(y,z,t) = \sum_{k} \sqrt{\lambda^p_k} a^p_k(t) \phi^p_k(y,z). 
\end{equation}
As mentioned above, the data set was symmmetrized, in order to compensate for the possible lack of statistical symmetry due to the presence of very long time scales in the dataset. 
However no significant difference was observed in the dominant modes corresponding to  the symmetrized and the unsymmetrized (original) dataset.  

POD analysis of the numerical pressure field
 was compared with the experimental results described in 
\cite{kn:bonnavion18} for an Ahmed body having the same base aspect ratio and at a Reynolds number of $Re=4\times10^5$.
The same symmetrization procedure was used for the experimental data.
The pressure was recovered from 21 sensors 
on the base of the body and then interpolated on a regular $10 \times 10$ grid. 
We note that the decomposition of the fluctuating pressure field yielded the same dominant modes 
at all ground heights investigated in the experiment for which bistability was observed and therefore we chose to show only results corresponding to the ground height $C=0.164H$.
We note that statistical symmetry was not generally observed in the experiments,
in particular since only a limited number (about 10) of switches was observed. 

\begin{figure}
\hspace{-2in}
\centerline{\includegraphics[height=7cm]{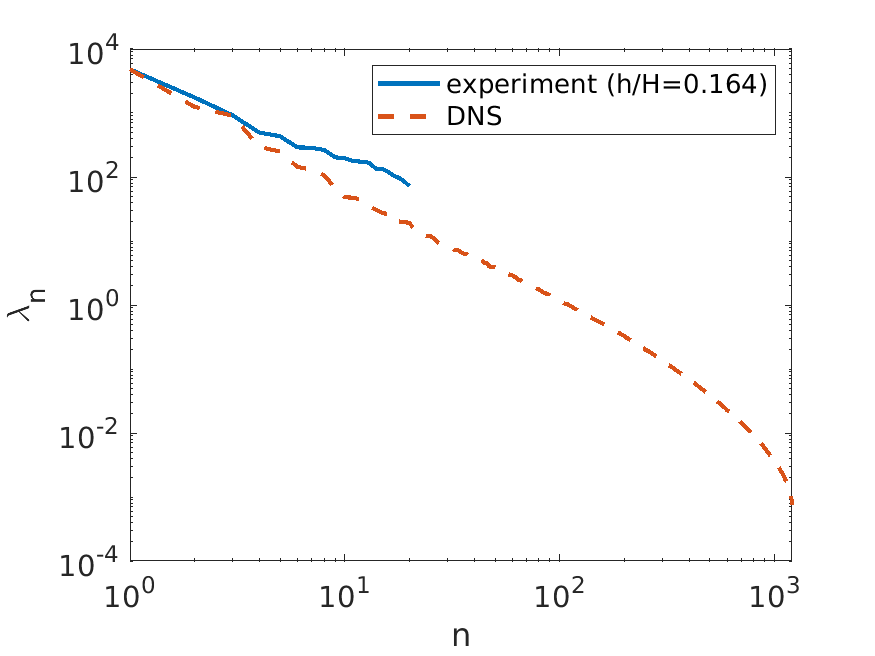}}
\caption{POD pressure spectrum in the simulation and in the experiment. \cite{kn:bonnavion18}.}
\label{eigenvalue}
\end{figure}

Figure ~\ref{eigenvalue} compares the experimental and the numerical POD 
spectrum. A good agreement is observed, given the discrepancy in the spatial resolution. 
Both spectra are characterized by a similar rate of decrease  $\lambda_n \sim n^{-1}$ for the largest-order modes. 
Based upon examination of the spectrum, we chose to focus on the 
first four modes, which capture 82 \% of the fluctuating energy in the simulation, with respective contributions of  55 \%, 14\%, 10\% and 3 \% of the 
fluctuations.

\begin{figure}
\hspace{-2in}
\begin{tabular}{rcc}
& \hspace{1.8in} Experiment & \hspace{1.8in} Simulation \\
\raisebox{0.5in}{n=1} & 
\includegraphics[height=4cm]{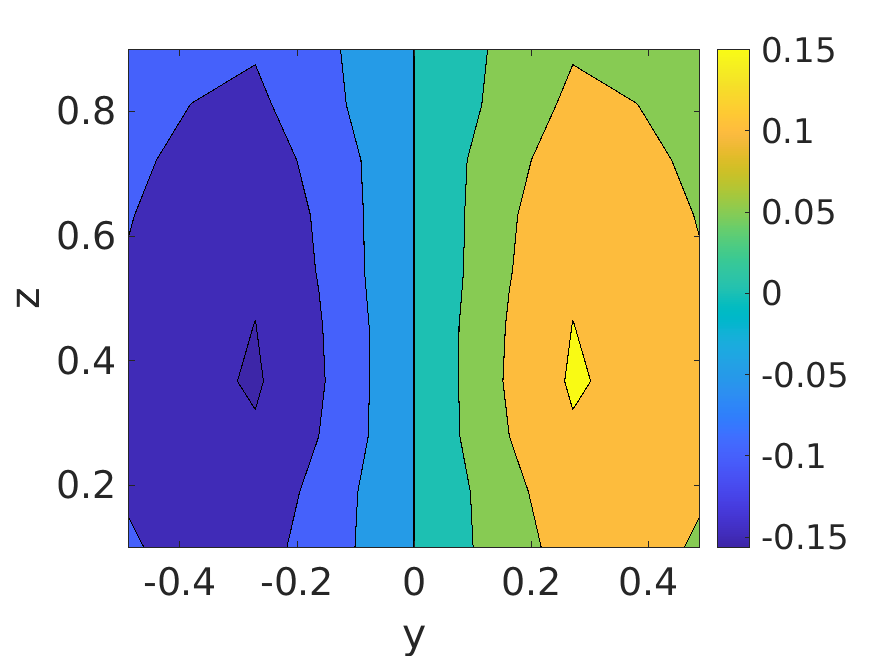} &
\includegraphics[height=4cm]{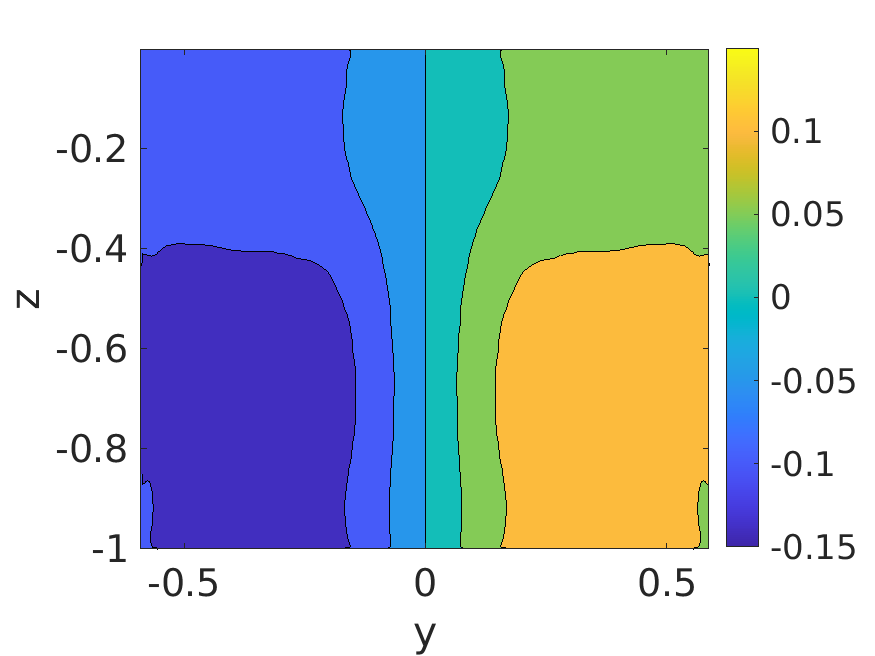} \\ 
\raisebox{0.75in}{n=2} & 
\includegraphics[height=4cm]{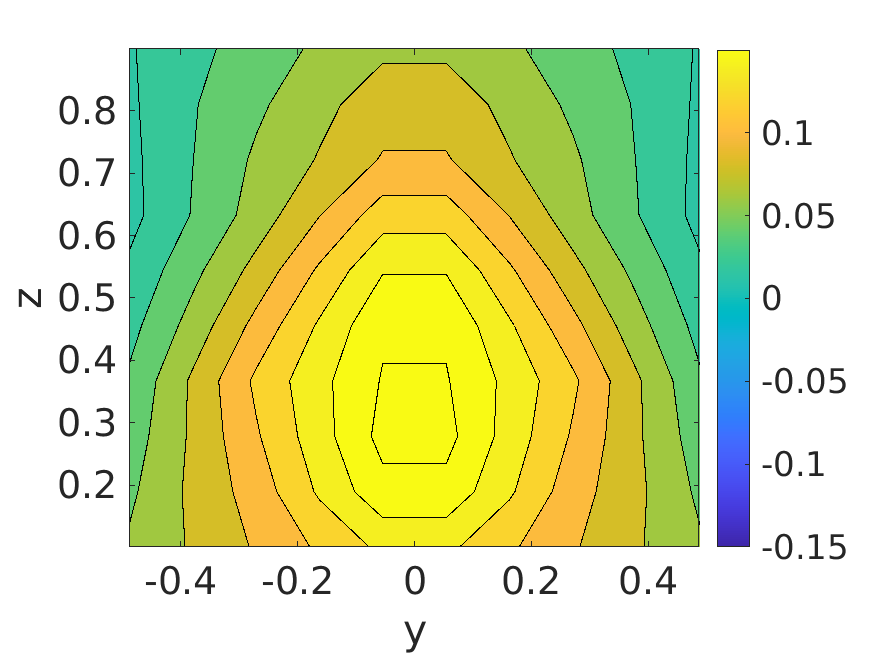} &
\includegraphics[height=4cm]{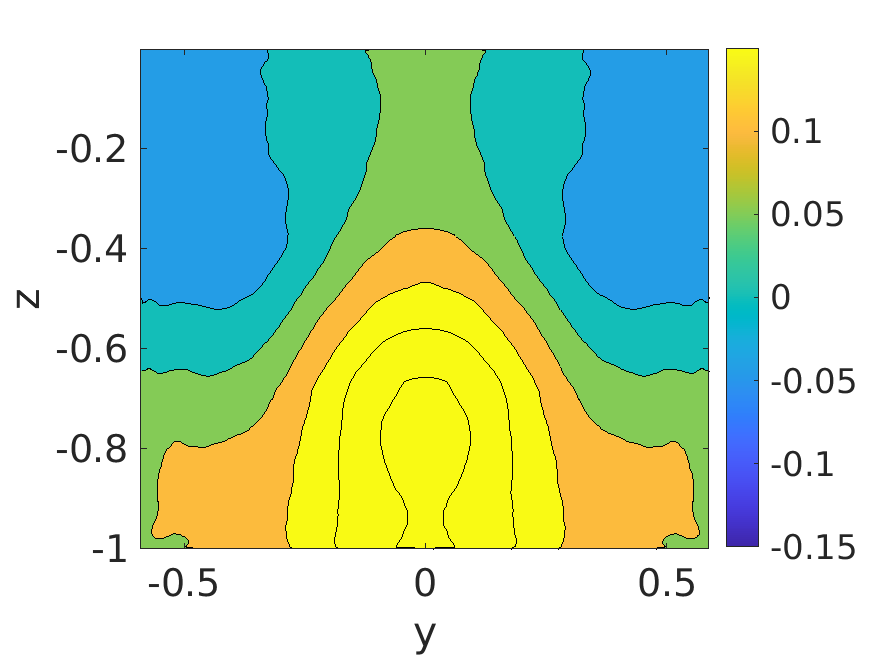} \\ 
\raisebox{0.75in}{n=3} & 
\includegraphics[height=4cm]{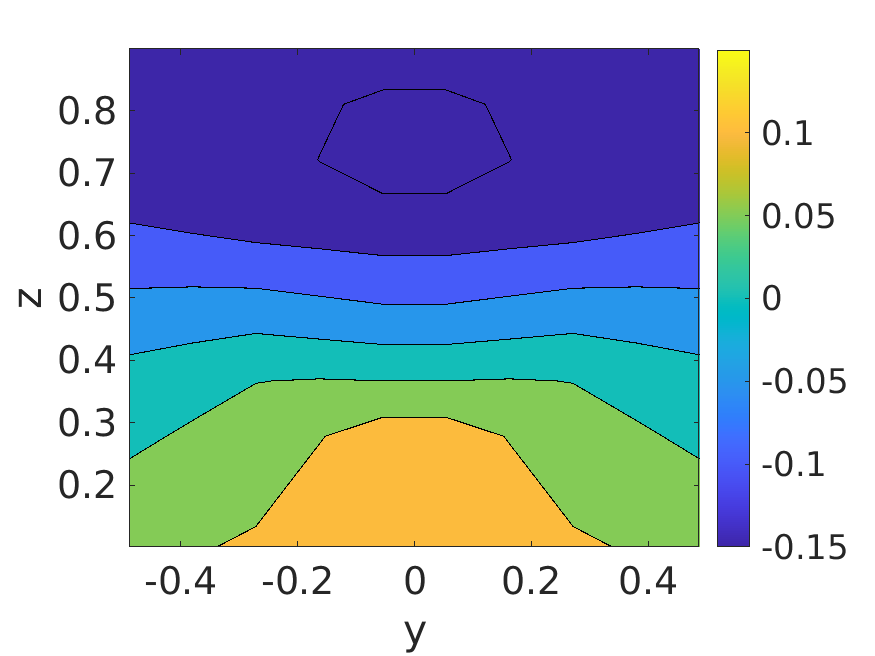} &
\includegraphics[height=4cm]{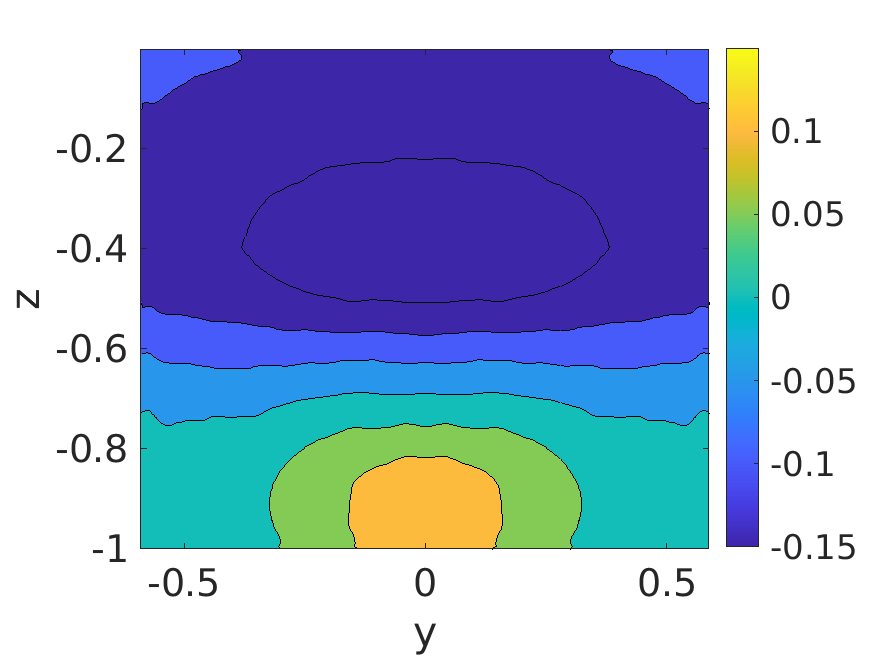} \\ 
\raisebox{0.75in}{n=4} & 
\includegraphics[height=4cm]{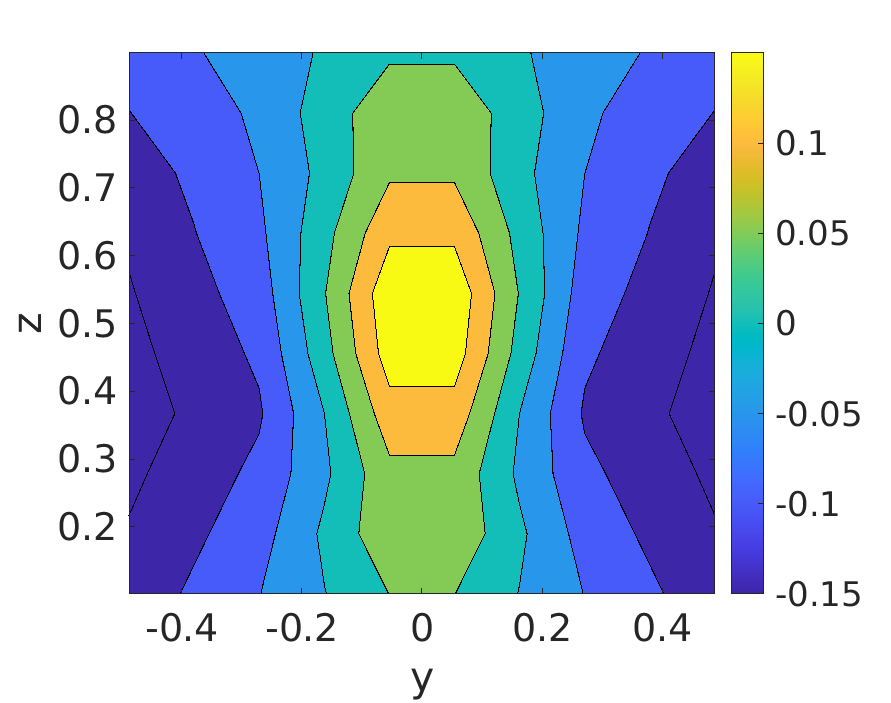} &
\includegraphics[height=4cm]{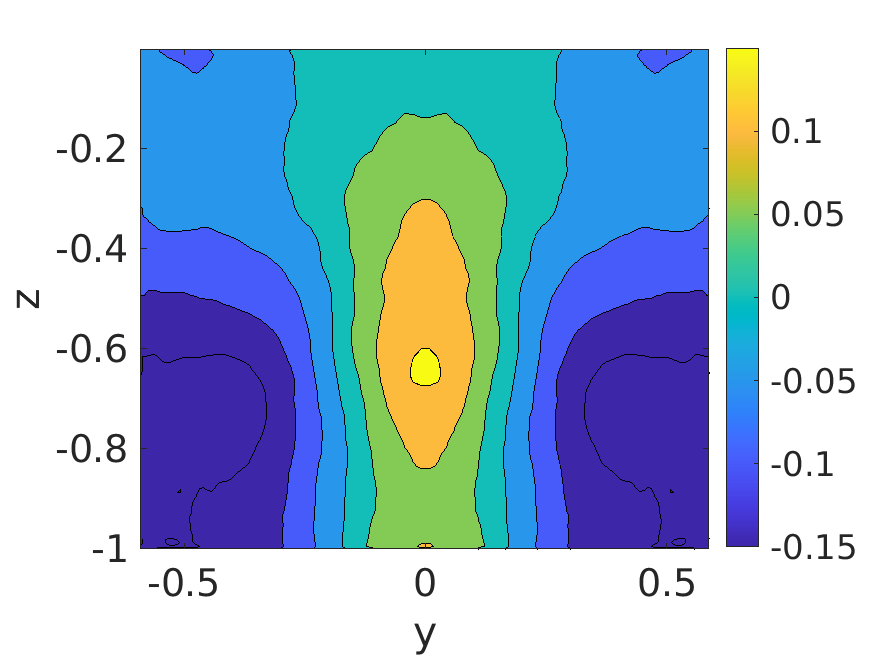} \\ 

\end{tabular}
\caption{POD pressure modes; Left: Experiment; Right: Simulation.   }
\label{podpressuremode}
\end{figure}

\begin{figure}
\hspace{-2in}
\begin{tabular}{ccc}

\includegraphics[height=3.4cm]{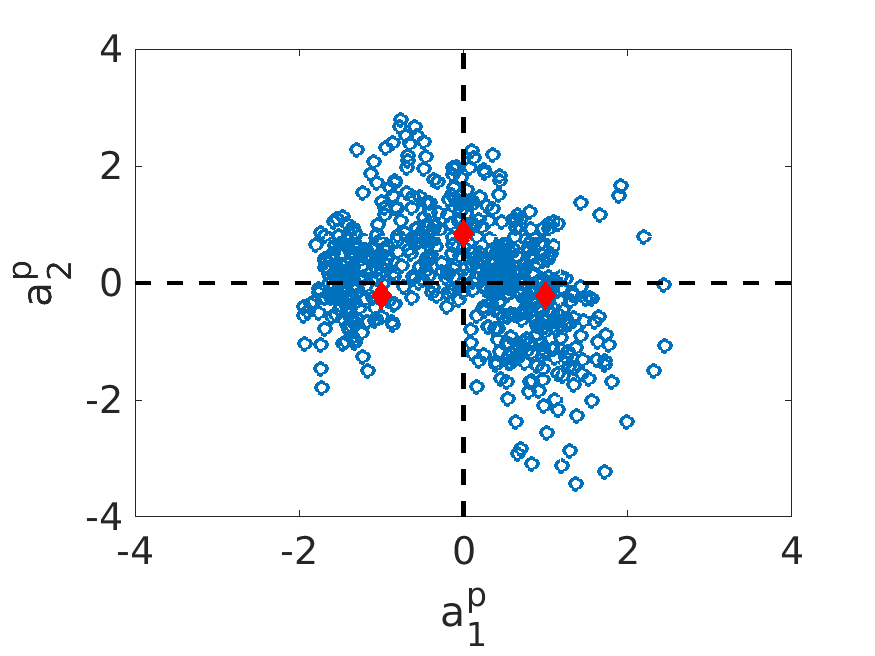}  &
\includegraphics[height=3.4cm]{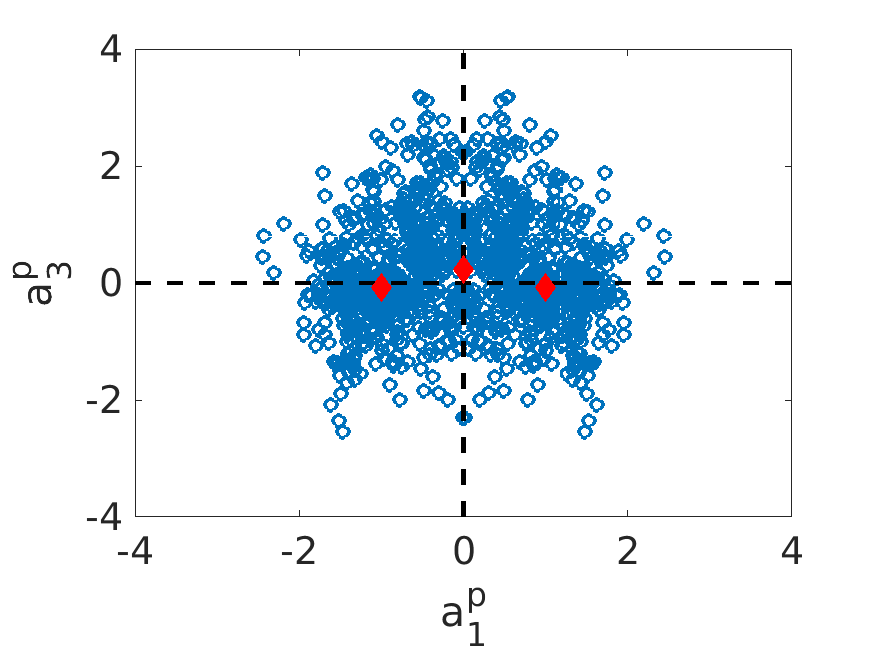} & 
\includegraphics[height=3.4cm]{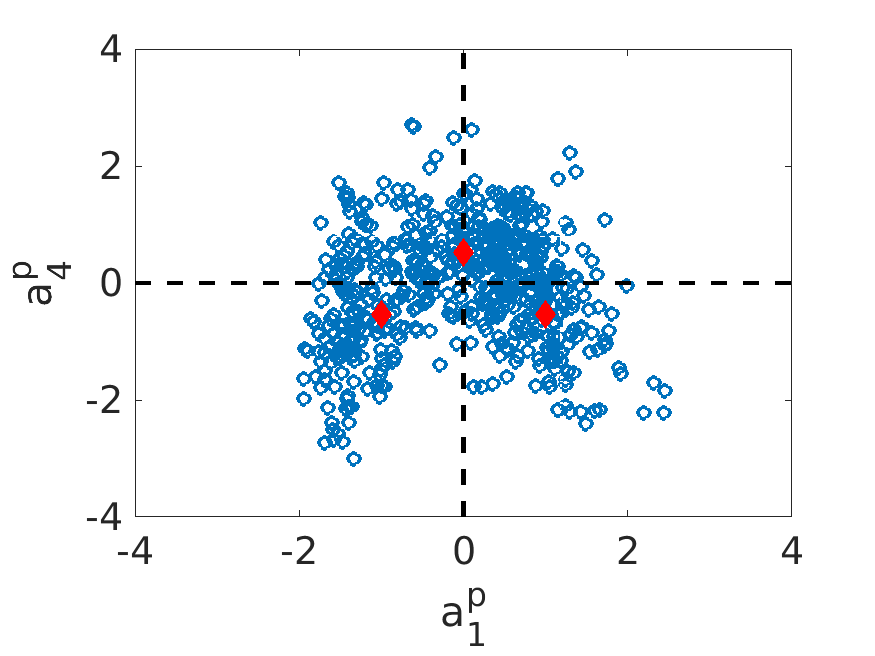}  \\
\includegraphics[height=3.4cm]{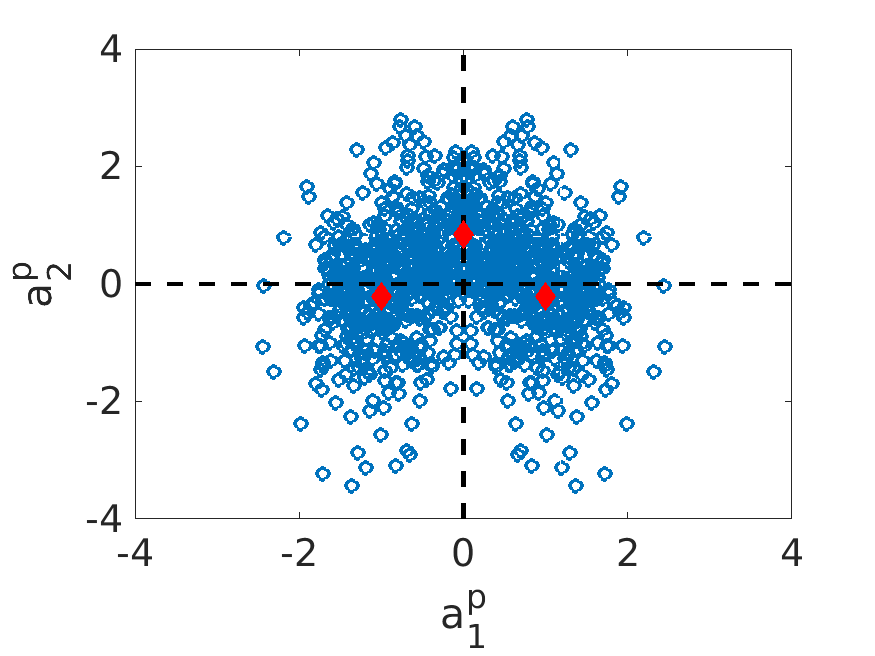}  &
\includegraphics[height=3.4cm]{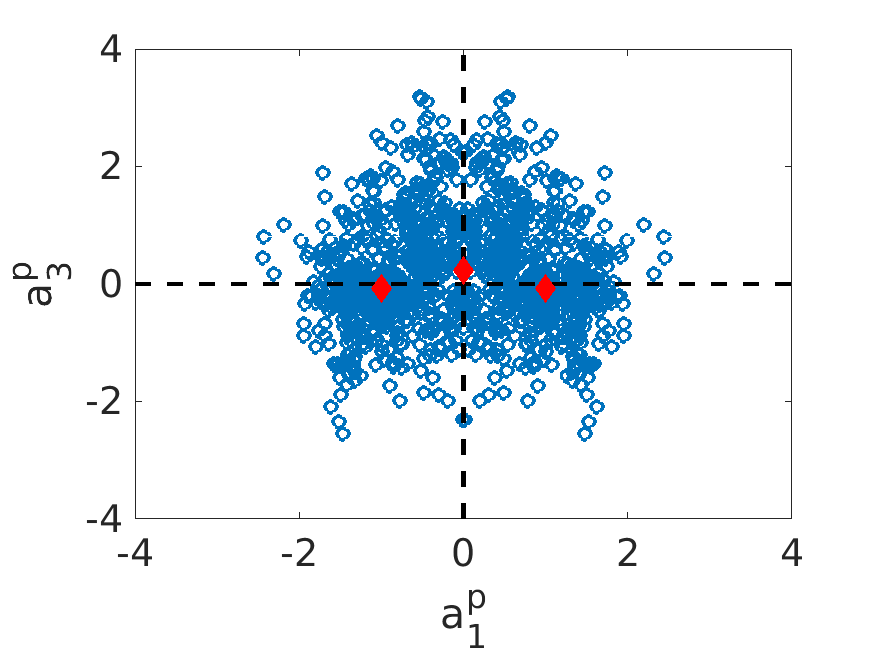} & 
\includegraphics[height=3.4cm]{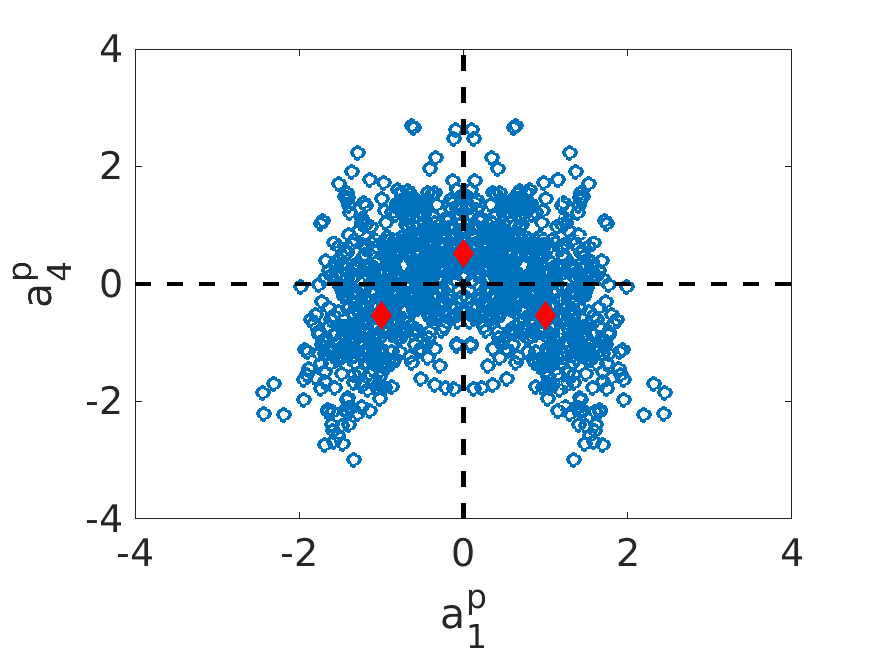}  \\
\includegraphics[height=3.4cm]{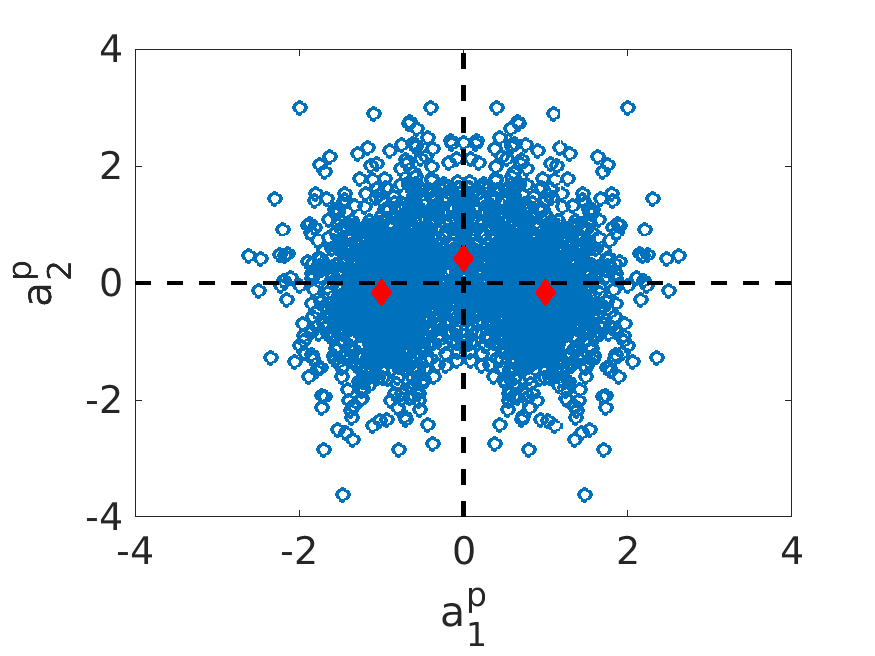}  &
\includegraphics[height=3.4cm]{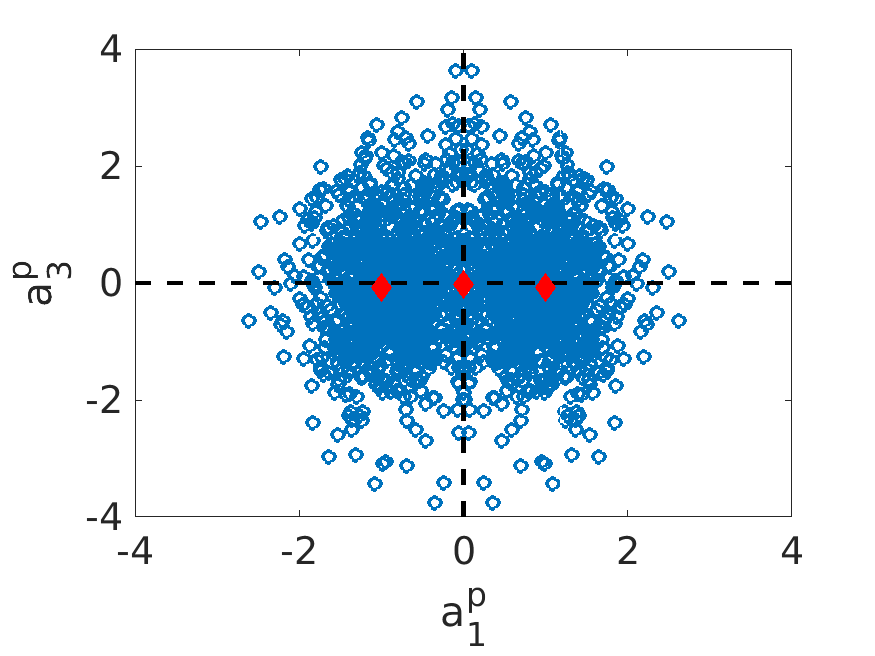}  &
\includegraphics[height=3.4cm]{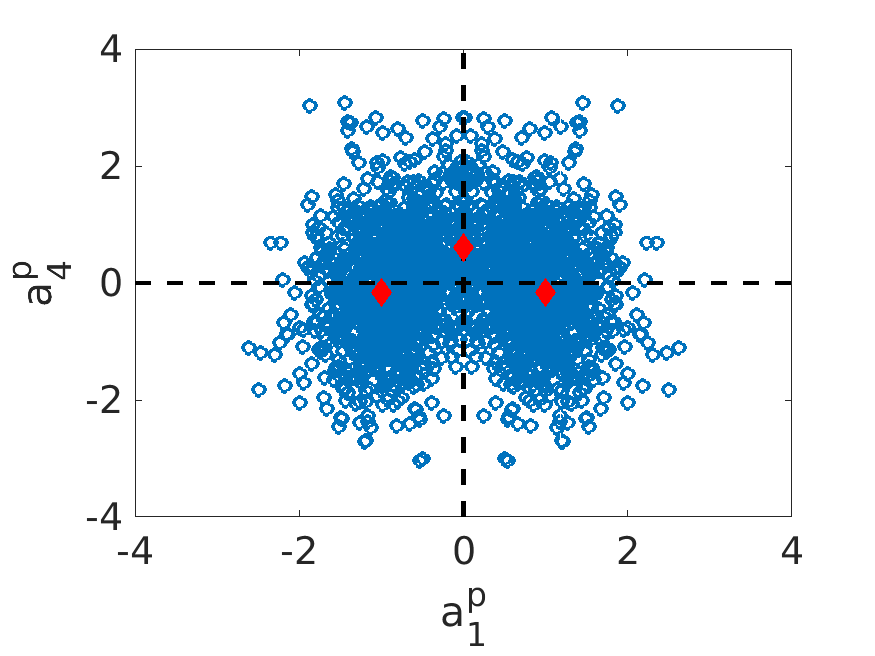}  \\
\end{tabular}
\caption{Phase portraits of the amplitudes in the simulation (top two rows)
and experiment (bottom row). From left to right: 
$(a_1^{p}, a_2^{p})$, $(a_1^{p}, a_3^{p})$
$(a_1^{p}, a_4^{p})$ in the simulation. The filled red diamonds correspond to the characteristic amplitudes $a_n^{p,eq}$ and
$a_n^{p,s}$.}
\label{phaseportrait}
\end{figure}

The spatial modes are represented in figure~\ref{podpressuremode} and their amplitude plotted in phase portraits in figure \ref{phaseportrait} for both the experiment \cite{kn:bonnavion18} and the simulation. The overall good agreement indicates that the large scale dynamics captured by these first POD modes are almost identical, basically made of switching between two wake deviations. 
The first fluctuating POD mode $n=1$ in figure~\ref{podpressuremode} is antisymmetric and corresponds to the well-identified wake deviation.
The sign of the mode amplitude corresponds to the orientation of the deviation.
When the flow is in a quasi-stationary state in figure~\ref{phaseportrait}, the normalized amplitude $a_1^p$ is close to $a_1^{p,eq}= \pm 1$. During a switching event, $a_1^p$ goes through a zero value, $a_1^{p,s}=0$. 
For $n \ge 2$, the three fluctuating modes are symmetric in figure~\ref{podpressuremode}.
Their characteristic amplitudes for the quasi-stationary state and the switching event are defined from conditional averages based on the deviation amplitude $a_1^{p}$:
$a_{n \ge 2}^{p,s}= E[a_n^p \vert \lvert a_1^p \rvert <0.1] $ and $ a_{n \ge 2}^{p,eq}= E[a_n^p \vert \lvert a_1^p \rvert >0.9].$  
They are spotted as red diamonds in phase portraits of figure~\ref{phaseportrait}. During the switching events, the three characteristic amplitudes $a_{n \ge 2}^{p,s}$ take either positive or null values.
The second mode $n=2$ in figure~\ref{podpressuremode} corresponds to a pressure variation of constant sign over the base. 
Since its amplitude takes positive values during the switches, then it is concluded that this second POD mode produces a global pressure increase during a switching event.
The third mode  $n=3$ is characterized by strong gradients in the vertical direction. Its characteristic amplitude is weakly negative for quasi-stationary states.
We note that the first three modes are similar to the pressure modes identified by
\cite{kn:pavia18} for the Windsor body (see figure 4 from their paper) and in particular display the same symmetries.
The fourth mode $n=4$ is dominated by vertical pressure gradients at the edges of the recirculation zone. Its characteristic amplitude is negative around quasi-stationary states and becomes positive during the switch, so that the pressure decreases on the lateral sides during the switch, which suggests that the local curvature is increased at the edges of the
recirculation zone during the switch.
A reconstruction based on the characteristic amplitudes identified
in figure \ref{phaseportrait} shows the evolution of the pressure
distribution as the flow goes through one switch. The pressure at the base globally increases by 5\%
during the switch mostly due to the second mode, indicating a drag reduction.
We note that the drag reduction level is comparable to the 7\% observed
by \cite{kn:pavia18} during a switch.

Figure~\ref{contributionpressure}($a$) shows the contribution of each mode to the variance of the base suction coefficient $c_b=-\frac{1}{HW}\iint_\text{base}c_p(y,z)dydz$, defined as $<c_{bn}'^2>/<c_{b}'^2>$ using the classical Reynolds decomposition $c_b(t)= C_b+ c_b'(t)$. The base suction, also called base drag is the base contribution to the drag coefficient. The variance of each POD mode
\begin{equation}
<c_{bn}'^2> =\lambda_n<{a^p_n}^2>\left(\frac{1}{HW}\iint_\text{base}\Phi^p_n(y,z)dydz \right)^2
\end{equation}
is simply related to the base suction variance by $<c_b'^2>=\sum_n <c_{bn}'^2>$. Only symmetric modes for which $\iint_\text{base}\Phi^p_n(y,z)dydz)^2\neq 0$ provide a non-zero contribution,
and it can be seen in figure~\ref{contributionpressure}($a$) that most of the base suction variations are due to the combined action of modes 2 and 3. 
Figure~\ref{contributionpressure}($b$) shows the average contribution to the base suction
coefficient for the characteristic events corresponding to a switch (s) or the wake in the quasi-stationary state (eq) identified earlier. Both contributions are evaluated with the conditional averaging $<c'_{bn}>_\text{eq}=\sqrt{\lambda_n}a_n^{p,\text{eq}}\frac{1}{HW}\iint_\text{base}\Phi^p_n(y,z)dydz$ and $<c'_{bn}>_\text{s}=\sqrt{\lambda_n}a_n^{p,\text{s}}\frac{1}{HW}\iint_\text{base}\Phi^p_n(y,z)dydz$. We can see that the most significant contribution is that of mode 2 during the switch, which is opposite to that of mode 3 and mode 4. 
The global effect of the first four POD modes is therefore to decrease the base suction during the switch.

\section{POD analysis of the velocity field in the near wake}
 We now consider the velocity field in the near-wake region, 
 taken here as $-0.5H < y < 0.5 H$, $-H < z < 0$, $3.8H < x < 5.5H$.
The modes are computed from the autocorrelation tensor limited to the near-wake region
defined above and extended to the full domain, as explained in the Appendix.
Figure \ref{velmodes} represents a view of the spatial modes in the horizontal mid-plane along with the power spectral density of 
the normalized amplitude $a_n^v$. The black lines show the limits of the POD
domain.

Since a POD analysis of the full velocity field was carried out in previous studies \citep{kn:podvin20,kn:podvin21}, we only provide a brief discussion of the modes.
The first two fluctuating modes respectively correspond to an antisymmetric, deviation mode and a symmetric, "switching" mode that
also experiences large variations during the switch.
The next-order modes are a mixture of vortex shedding dynamics (as can be seen particularly for modes 4, 6, 8) and characterized by a peak at the  
frequency of $fH/U_\infty \sim 0.2$.  Most modes are characterized by strong shear layers at the edges of the recirculation zone.

\section{POD-based linear stochastic estimation}

\subsection{Method}

In this section we describe the principle of the POD-LSE method \cite{kn:expif18}.
We suppose that two different quantities $\qo$ and $\qt$ can be extracted simultaneously from a set of $N$ snapshots.
The quantities may have different dimensions (in particular be scalars) and can be defined on the same or different
parts of the domain.
Independent POD decomposition of each set of snapshots  yields 
\begin{equation} 
\qq(\xx,t_i)=\sum_{n=1}^{N} a_n^{\qq}(t_i) \pphi_n^{\qq}(\xx),~~\qo(\xx,t_i)=\sum_{n=1}^{N} a_n^{\qo}(t_i) \pphi_n^{\qo}(\xx).
\end{equation} 

The idea is then to use POD to estimate one quantity (for instance $\qo$) based on measurements of the other (for instance $\qt$).
For each time $t_i$, $i=1, \ldots N$,   the $N$ coefficients $a_n^{\qo}(t_i)$ can be obtained from the mapping 
\begin{equation}
a_n^{\qo}(t_i)=M_{nk}^{\qo\qt} a_{k}^{\qt}(t_i) 
\end{equation}
where \[ M^{\qo\qt}=A^{\qo} (A^{\qt})^T \]
$A^{\qo }$ and $A^{\qt }$ are respectively the matrices consisting of  rows of coefficients $ a_k^{\qo}(t_i)$ and $a_k^{\qo}(t_i)$:
\[ A^{\qo}= \begin{bmatrix} a_1^{\qo}(t_1) & a_1^{\qo}(t_2) & \ldots  & a_1^{\qo}(t_N)\\
a_2^{\qo}(t_1) & a_2^{\qo}(t_2) & \ldots  & a_2^{\qo}(t_N)\\
& & \ldots & \\
a_N^{\qo}(t_1) & a_N^{\qo}(t_2) & \ldots & a_1^{\qo}(t_N) \\
.
\end{bmatrix}
\]

The expression is exact for the snapshots of the database $t_i$ if the full $N-th$ order matrix is used.
However if the matrix is truncated, one can still obtain 
an estimate at any time $t$ of 
the first $N_{\qo} \leq N$ POD amplitudes based on a subset of $N_{\qt} \leq N$ POD amplitudes using 
\begin{equation}
a_n^{\qo e}(t)=\tM_{nk}^{\qo\qt} a_k^{\qt}(t) 
\label{LSE}
\end{equation}
where $\tM$ is a sub-matrix of $M$ containing the $N_{\qo}$ rows and $N_{\qt}$ columns.
A high value of the matrix entry  $\tM_{nk}^{\qo\qt}$  indicates that the $k$-th mode of $\qt$  has a strong influence on 
the $n$-th mode of $\qo$.
The final step of the estimation method is to normalize the estimate
with its standard deviation on the set of POD snapshots:   
\[ a_k^{\qo e}(t) \rightarrow \frac{\sqrt{N}}{\sqrt{\sum_{i=1}^{N} \lvert a_k^{\qo e}(t_i) \rvert^2}}   a_k^{\qo e}(t)   \]

In the paper the method is applied to the base pressure $p$ 
and the velocity field $\uu$ in the near-wake.
The aim is to determine whether a linear relationship exists between 
the base pressure modes and the velocity modes.

\subsection{Determining the velocity field from the pressure}

We first investigate the dependence of the velocity field on the base pressure in the near wake. We then examine the largest velocity modes, focusing on the large scales with a moving time average 
of 5 convective time units, which corresponds to the vortex shedding period.
Figure \ref{estimvel12} and \ref{estimvel910} compare the evolution of the exact velocity mode amplitude with 
its pressure-based estimation using 
\begin{equation}
a_n^{v e}(t)=\tM_{nk}^{vp} a_k^{p}(t) 
\end{equation}
We can see that the first two POD velocity modes, i.e the deviation and the switch mode,  are well estimated from the pressure
coefficients. 
Figure \ref{estimvel12} shows that the first fluctuating velocity and pressure modes are perfectly correlated,
while the switch mode can be estimated from the difference from the second and third  POD pressure modes. 
Table \ref{coefcorrelvel} presents the correlation coefficient between the velocity POD amplitudes and the pressure-based estimates.
We can see that modes 8, 9 and  10 are the best correlated with the pressure estimation, in particular if a moving average of
length 5 time units is applied in order to remove the high frequencies 
associated with vortex shedding. 
Figure \ref{estimvel910} shows that the POD amplitude of modes 9 and 10 is well estimated.  
Figure \ref{velrecons} compares instantaneous fields with their projection on the ten most energetic velocity modes and estimation
for three different times (indicated by the vertical lines in figures \ref{estimvel12} and \ref{estimvel910}) corresponding to a quasi-stationary asymmetric state,
a nearly symmetric state, and the opposite asymmetric state. 
We can see that the large scales of the flow are very well estimated from the pressure base measurements.

\subsection{Reconstructing  the  pressure from the near-wake field}

Figure \ref{estimpres} compares the  real pressure POD amplitudes (extracted from the simulation) 
$ a_n^p$ with their estimation
$a_n^{p e}$,  based on  10 velocity modes using equation (\ref{LSE}):
\begin{equation}
a_n^{p e}(t)=\tM_{nk}^{pv} a_k^{v}(t).
\end{equation}
A good agreement is observed,  
with a nearly perfect correlation for the first fluctuating mode corresponding
to the deviation amplitude,  and high correlation coefficients particularly for the second and the third modes as can be seen in table 
~\ref{coefcorrelpres}.   
We note that due to antisymmetry, the dominant deviation pressure mode does not contribute to the base suction fluctuation $c'_b(t)$, the variations
of which are due to symmetric modes only. 

Figure \ref{Mpv23} shows that the second pressure mode is mostly determined by velocity modes 2, 3, and 9 
while the third pressure mode was mostly associated with mode 2 and 10.
This is not entirely surprising since  the imprint of the  pressure modes
on velocity modes 9 and 10 was found to be large. 
Since the pressure modes 2 and 3 contribute most to the pressure coefficient (figure \ref{contributionpressure}), we constructed an estimation of  the pressure coefficient from the second and third POD pressure modes
\begin{equation}
c_b'^{\text{e}}=-\sum_{n=2}^{3} \lambda_n^{p,1/2} a_n^{p e} \frac{1}{HW}\iint_\text{base} \phi_n^p dydz.
\end{equation}
which we compared with the equivalent projection
\begin{equation}
c_b'^{\text{proj}}=-\sum_{n=2}^{3} \lambda_n^{p,1/2} a_n^{p} \frac{1}{HW}\iint_\text{base}\phi_n^p dydz.
\end{equation}

Results are shown in figure  \ref{dragcoef}. As expected, the agreement between the full pressure coefficient and its projection is excellent. A very good agreement 
is obtained for the estimation.
The correlation coefficient between the full drag and estimated one  is  0.7 and increases to 0.8 with a moving average of length 5 convective units.

\section{Conclusion}

We have investigated how  the wake dynamics in the 
flow around an Ahmed body can be described using POD analysis of the base pressure.
Decomposition of the pressure field  shows that 
the switch is characterized by the following modifications of the 
pressure distribution:
i) a global increase over the body base ii) a gradient in the vertical direction
and iii) a symmetric lateral gradient within the 
recirculation zone corresponding to a pressure 
increase along the mid-vertical plane and decrease in the 
outer recirculation zone. 
These features were identified in both the numerical simulation 
and in experimental results.
The signature of the pressure modes could clearly be identified in the evolution of
the dominant POD velocity modes.  
The fluctuating velocity field, in particular the most energetic deviation and switch modes, was 
well recovered from pressure measurements.
Conversely, variations of the pressure drag coefficient, which are essentially determined by the largest two symmetric
pressure modes, could be well recovered from the near-wake velocity field.

\begin{appendix}
\section{POD}

The main tool of analysis used in this paper is Proper
Orthogonal Decomposition (POD) \cite{kn:lumleyPOD}.
We consider a spatio-temporal vector field $\tilde{\qq}(\xx,t)$
defined on a spatial domain $D$.
$\tilde{\qq}(\xx,t)$ will refer either to the pressure field $p$ or the velocity field $\uu$.
The fluctuating part of the field $\qq$ 
(with respect to its temporal mean) an be expressed 
as a superposition of spatial modes
\begin{equation} 
\qq(\xx,t) = \sum_{k} \tilde{a}_k(t) \pphi_k(\xx) 
\end{equation}
where the spatial modes $\pphi_k$ are orthogonal (and can be made orthonormal), i.e
\[ \int_D \pphi_k(\xx) . \pphi_m(\xx) d\xx = \delta_{km}, \]
and the amplitudes $a_k$ are uncorrelated.
The modes  can be ordered by decreasing energy $\lambda_1 \ge \lambda_2 \ge \ldots \ge \lambda_k = <\tilde{a}_k \tilde{a}_k >$, where
$<.>$ represents a time average.

The amplitudes $\tilde{a}_k$ can be obtained from the knowledge of the spatial modes by projection of the vector field $\qq$
onto the spatial modes 
\begin{equation} 
\tilde{a}_k(t)= \int_{\Omega} \uu(\xx,t). \pphi_k(\xx) d\xx.  
\end{equation}
%

The modes $\phi_k$ and the values $\lambda_k$ can be obtained from the eigenproblem (direct method) 
\begin{equation}
\int_{D} C(\xx,\xx') . \pphi_m^q(\xx') d\xx' = \lambda_{k} \pphi_k(\xx) 
\end{equation}
where $C$ is the spatial autocorrelation tensor
\[ C(\xx,\xx')=  < \qq(\xx,t) \qq(\xx',t') >  \]

Alternatively, if the number of snapshots $N$ used to compute $C$ is smaller than the spatial dimension
of the problem, one can obtain the modes through the method of snapshots where 
\begin{equation}
\bar{C}_{ij}. A_{jk} = \lambda_{k} A_{ik} 
\end{equation}

where $A_{jk}=a_k(t_j)$
and $\bar{C}$ is the temporal autocorrelation matrix 
\[ \bar{C}_{ij}=\frac{1}{N} \int_D \qq(\xx,t_i) . \qq(\xx, t_j) d \xx \]

We note that spatial modes obtained for quantity $\qq$ on a domain $D$ can be extended to another 
quantity $\qq'$ (which can be for instance the same quantity $\qq$ on a domain $D'$)
\cite{kn:boree} using
\begin{equation}
\phi_n^{q'}(\xx,t)=\sum_{k=1}^{N} a_n^q(t_k) \qq'(\xx,t_k)
\end{equation}

In the paper we will consider pressure and velocity decompositions
and will index the corresponding amplitudes as respectively
$a_k^p$ and $a_k^v$.
In the paper  we will consider normalized amplitudes defined as
$a_k= \tilde{a}_k/\sqrt{\lambda_k}.$

\end{appendix}

\pagebreak


\begin{thebibliography}{10}

\bibitem{kn:mo&adrian}
R.J. Adrian and P.~Moin.
\newblock Stochastic estimation of organized turbulent structure : homogeneous
  shear flow.
\newblock {\em J. Fluid Mech.}, 190:531--559, 1988.

\bibitem{kn:barros14}
D.~Barros, T.~Ruiz, J.~Borée, and B.N. Noack.
\newblock Control of three-dimensional blunt body wake using low and high
  frequency pulsed jets.
\newblock {\em International Journal of Flow Control}, 6(1):61--74, 2014.

\bibitem{kn:bonnavion18}
G.~Bonnavion and O.~Cadot.
\newblock Unstable wake dynamics of rectangular flat-backed bluff bodies with
  inclination and groud proximity.
\newblock {\em J. Fluid Mech.}, 854:196--232, 2018.

\bibitem{kn:bonnet94}
J.P. Bonnet, D.R. Cole, J.~Delville, M.N. Glauser, and L.S. Ukeiley.
\newblock Stochastic estimation and proper orthogonal decomposition:
  Complementary techniques for identifying structure.
\newblock {\em Exp. Fluids}, 17:307--314, 1994.

\bibitem{kn:boree}
J.~Bor\'ee.
\newblock Extended proper orthogonal decomposition: A tool to analyse
  correlated events in turbulent flows.
\newblock {\em Exp. in Fluids}, 2003.

\bibitem{kn:brackston16}
R.D. Brackston, J.M. Garci De~La Cruz, A.~Wynn, G.~Rigas, and J.F. Morrison.
\newblock Stochastic modelling and feedback control of bistability in a
  turbulent bluff bodywake.
\newblock {\em J. Fluid Mech.}, 802:726--749, 2016.

\bibitem{kn:chang99}
P.A. {Chang III}, U.~Piomelli, and W.K. Blake.
\newblock Relationship between wall pressure and velocity-field sources.
\newblock {\em Phys. Fluids}, 11:3434, 1999.

\bibitem{kn:citriniti2000}
J.~Citriniti and W.~George.
\newblock Reconstruction of the global velocity field in the axisymmetric
  mixing layer utilizing the proper orthogonal decomposition.
\newblock {\em J. Fluid Mech.}, 418:137--166, 2000.

\bibitem{kn:dallalonga19}
L.~Dalla~Longa, O.~Evstafyeva, and A.~S. Morgans.
\newblock Simulations of the bi-modal wake past three-dimensional blunt bluff
  bodies.
\newblock {\em Journal of Fluid Mechanics}, 866:791–809, 2019.

\bibitem{kn:durgesh10}
V.~Durgesh and J.W. Naughton.
\newblock Multi-time delay, multi-point linear stochastic estimation of a
  cavity shear layer velocity from wall-pressure measurements.
\newblock {\em Experiments in Fluids}, 49:571--583, 2010.

\bibitem{kn:evrard16}
A.~Evrard, O.~Cadot, V.~Herbert, D.~Ricot, R.~Vigneron, and J.~Delery.
\newblock Fluid force and symmetry breaking modes of a 3d bluff body with a
  base cavity.
\newblock {\em Journal of Fluids and Structures}, 61:99--114, 2016.

\bibitem{kn:evstafyeva17}
O.~Evstafyeva, A.~Morgans, and L.~Dalla Longa.
\newblock Simulation and feedback control of the {Ahmed} body flow exhibiting
  symmetry breaking behaviour.
\newblock {\em J. Fluid Mech.}, 817, 2017.

\bibitem{kn:fan20}
Y.~Fan, X.~Chao, S.~Chu, Z.~Yang, and O.~Cadot.
\newblock Experimental and numerical analysis of the bi-stable turbulent wake
  of a rectangular flat-backed bluff body.
\newblock {\em Phys. Fluids}, 32:105111, 2020.

\bibitem{kn:ferreira21}
M.A. Ferreira and B.~Ganapathisubramani.
\newblock Scale interactions in velocity and pressure within a turbulent
  boundary layer developing over a staggered-cube array.
\newblock {\em J. Fluid Mech.}, 910:A48, 2021.

\bibitem{kn:goda79}
K.~Goda.
\newblock A multistep technique with implicit difference schemes for
  calculating two- or three-dimensional cavity flows.
\newblock {\em J. Comp. Phys.}, 30:76--95, 1979.

\bibitem{kn:grandemange13}
M.~Grandemange, M.~Gohlke, and O.~Cadot.
\newblock Turbulent wake past a three-dimensional blunt body. part 1. global
  modes and bi-stability.
\newblock {\em Journal of Fluid Mechanics}, 722:51–84, 2013.

\bibitem{kn:hudy07}
L.~Hudy and A.~Naguib.
\newblock Stochastic estimation of a separated-flow field using
  wall-pressure-array measurements.
\newblock {\em Phys. Fluids}, 19:024103, 2007.

\bibitem{kn:lasagna13}
D.~Lasagna, M.~Orazi, and G.~Iuso.
\newblock Multi-time delay, multi-point linear stochastic estimation of a
  cavity shear layer velocity from wall-pressure measurements.
\newblock {\em Phys. Fluids}, 25:017101, 2013.

\bibitem{kn:li16}
R.~Li, D.~Barros, J.~Bor\'ee, O.~Cadot, B.R. Noack, and L.~Cordier.
\newblock Feedback control of bimodal wake dynamics.
\newblock {\em Exp Fluid}, 51(4):158, 2016.

\bibitem{kn:lucas17}
J.M. Lucas, O.~Cadot, V.~Herbert, S.~Parpais, and J.~D\'elery.
\newblock A numerical investigation of the asymmetric wake mode of a squareback
  ahmed body - effect of a base cavity.
\newblock {\em J. Fluid Mech.}, 831:675--697, 2017.

\bibitem{kn:lumleyPOD}
J.L. Lumley.
\newblock The structure of inhomogeneous turbulent flows.
\newblock In A.M Iaglom and V.I Tatarski, editors, {\em Atmospheric Turbulence
  and Radio Wave Propagation}, pages 221--227. Nauka, Moscow, 1967.

\bibitem{kn:pavia18}
G.~Pavia, M.~Passmore, and C.~Sardu.
\newblock Evolution of the bi-stable wake of a square-back automotive shape.
\newblock {\em Exp. in Fluids}, 59:2742, 2018.

\bibitem{kn:perry16}
A.K. Perry, G.~Pavia, and M.~Passmore.
\newblock Influence of short rear end tapers on the wake of a simplified
  square-back vehicle: wake topology and rear drag.
\newblock {\em Experiments in Fluids}, 57(11), 2016.

\bibitem{kn:plumejeau19}
B.~Plumejeau, S.~Delpart, L.~Keirsbulck, M.~Lippert, and W.~Abassi.
\newblock Ultra-local model-based control of the square-back ahmed body wake
  flow.
\newblock {\em Phys. Fluids}, 31:085103, 2019.

\bibitem{kn:expif18}
B.~Podvin, S.~Nguimatsia, J.M. Foucaut, C.~Cuvier, and Y.~Fraigneau.
\newblock On combining linear stochastic estimation and proper orthogonal
  decomposition for flow reconstruction.
\newblock {\em Experiments in Fluids}, 59(3):58, 2018.

\bibitem{kn:podvin21}
B.~Podvin, S.~Pellerin, Y.~Fraigneau, G.~Bonnavion, and O.~Cadot.
\newblock Low-order modelling of the wake dynamics of an ahmed body.
\newblock {\em J. Fluid Mechanics}, 927:R6, 2021.

\bibitem{kn:podvin20}
B.~Podvin, S.~Pellerin, Y.~Fraigneau, A.~Evrard, and O.~Cadot.
\newblock Proper orthogonal decomposition analysis and modelling of the wake
  deviation behind a squareback ahmed body.
\newblock {\em Phys. Rev. Fluids.}, 6(5):064612, 2020.

\bibitem{kn:rao18}
A.~Rao, G.~Minelli, B.~Basara, and S.~Krajnovic.
\newblock On the two flow states in the wake of a hatchback ahmed body.
\newblock {\em J. Wind Eng. Ind. Aerodyn.}, 173:262--278, 2018.

\bibitem{kn:siro87}
L.~Sirovich.
\newblock Turbulence and the dynamics of coherent structures part i: Coherent
  structures.
\newblock {\em Quart. Appl. Math.}, 45(3):561--571, 1987.

\bibitem{kn:taylorglauser02}
J.~Taylor and M.~Glauser.
\newblock Toward practical flow sensing and control via pod and lse-based
  low-dimensional tools.
\newblock {\em ASME 2002 Fluids Engineering Division Summer Meeting, Montreal,
  Quebec, 14-18 July}, 2002.

\bibitem{kn:tinney07}
C.E. Tinney, F.~Coiffet, J.~Delville, A.M. Hall, P.~Jordan, and M.N. Glauser.
\newblock On spectral linear stochastic estimation.
\newblock {\em Experiments in Fluids}, 41:763--775, 2007.

\end{thebibliography}

\pagebreak
%
%

%
%

%
%

%
%

%
%
\begin{figure}
\hspace{-2in}
\begin{tabular}{ccc}
\includegraphics[height=3.4cm]{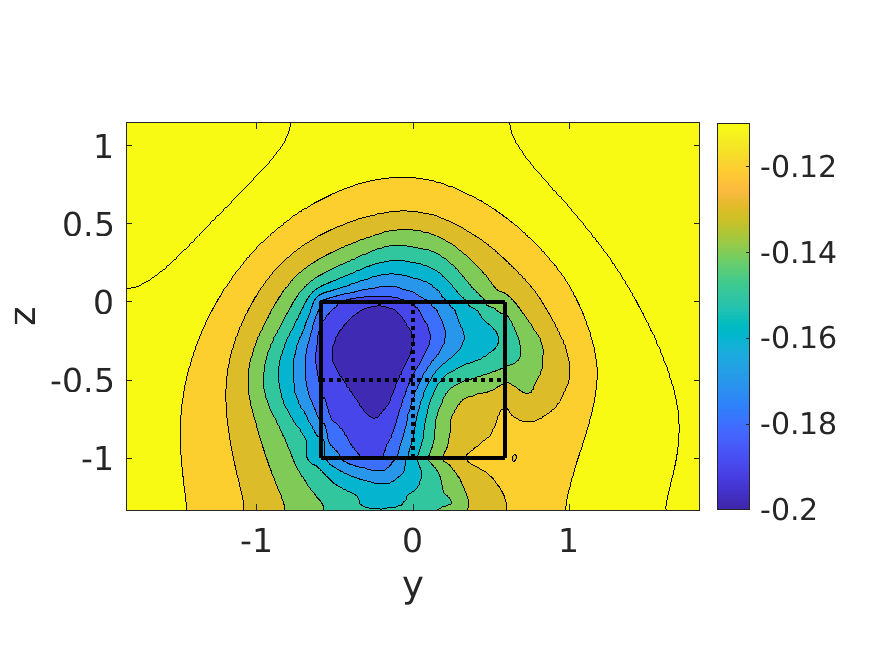} &
\includegraphics[height=3.4cm]{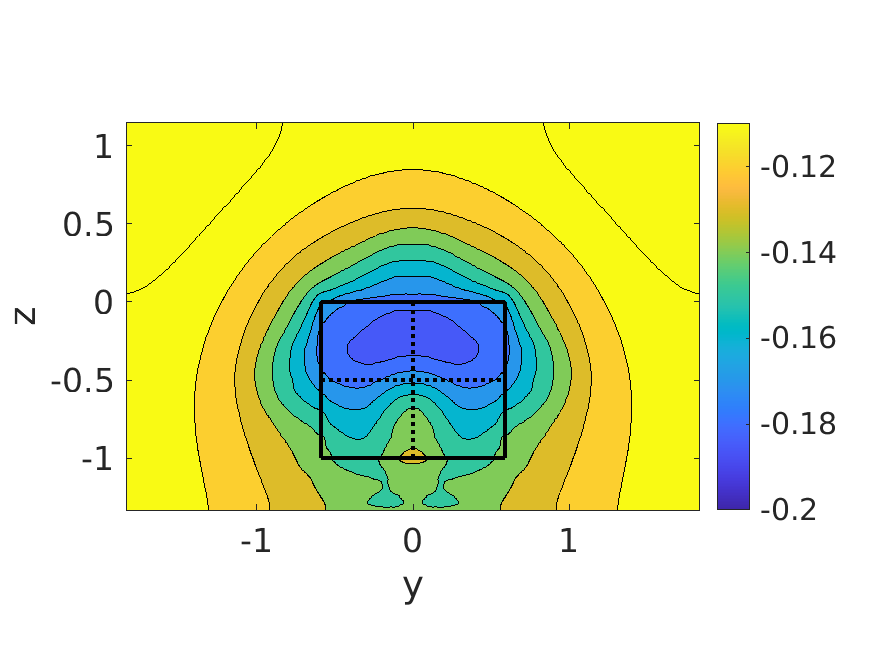} &
\includegraphics[height=3.4cm]{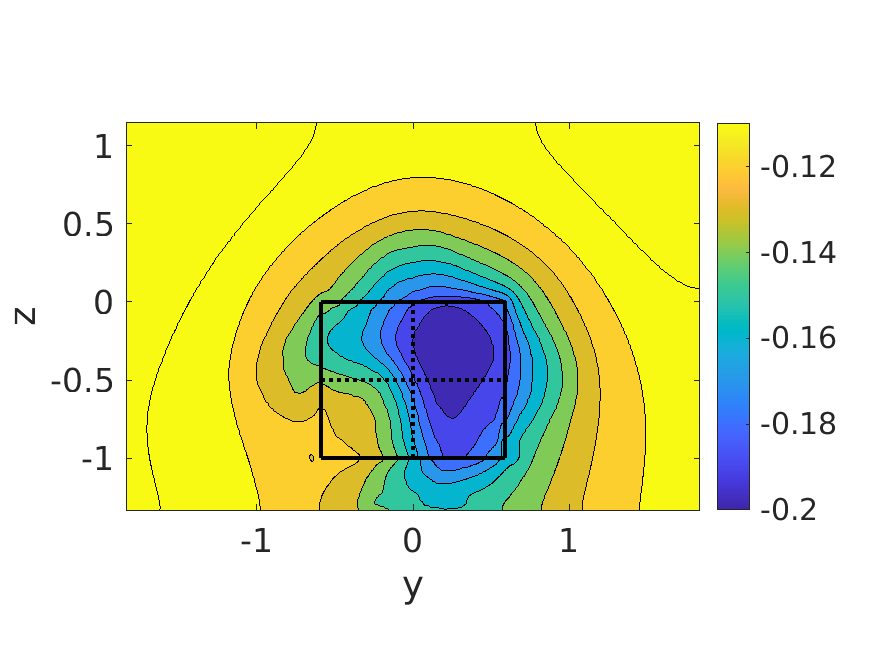} \\ 
\end{tabular}
\caption{Reconstruction of the mean pressure coefficient distribution at the base $c_p(y,z)$ using the symmetrized mean
and the first four POD modes
based on the characteristic amplitudes $a_n^{p,eq}$ and $a_n^{p,s}$
left) quasistationary state  $C_p + \sqrt{\lambda_1^p}  \phi_1^p + \sum_{n=2}^{4} \sqrt{\lambda_n^p} a_n^{p,eq} \phi_n^p$;
center) switch state $C_p + \sum_{n=2}^{4} \sqrt{\lambda_n^p} a_n^{p,s} \phi_n^p$;
right) symmetric quasistationary state $C_p - \sqrt{\lambda_1^p}  \phi_1^p + \sum_{n=2}^{4} \sqrt{\lambda_n^p} a_n^{p,eq} \phi_n^p$.}
\label{presrecons} 
\end{figure}

\begin{figure}[h]
\hspace{-2in}
\begin{tabular}{cc}
\includegraphics[height=5cm]{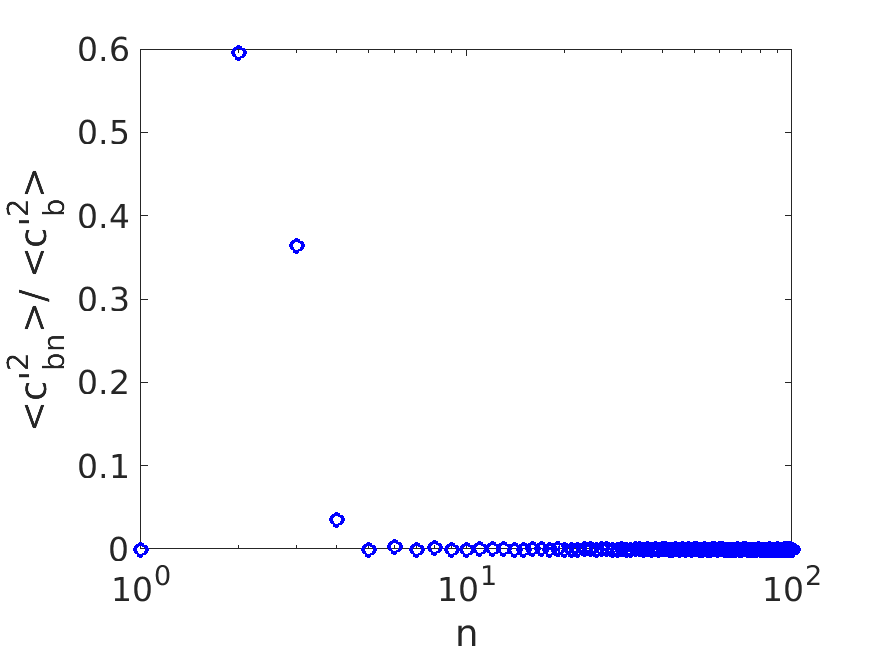} &
\includegraphics[height=5cm]{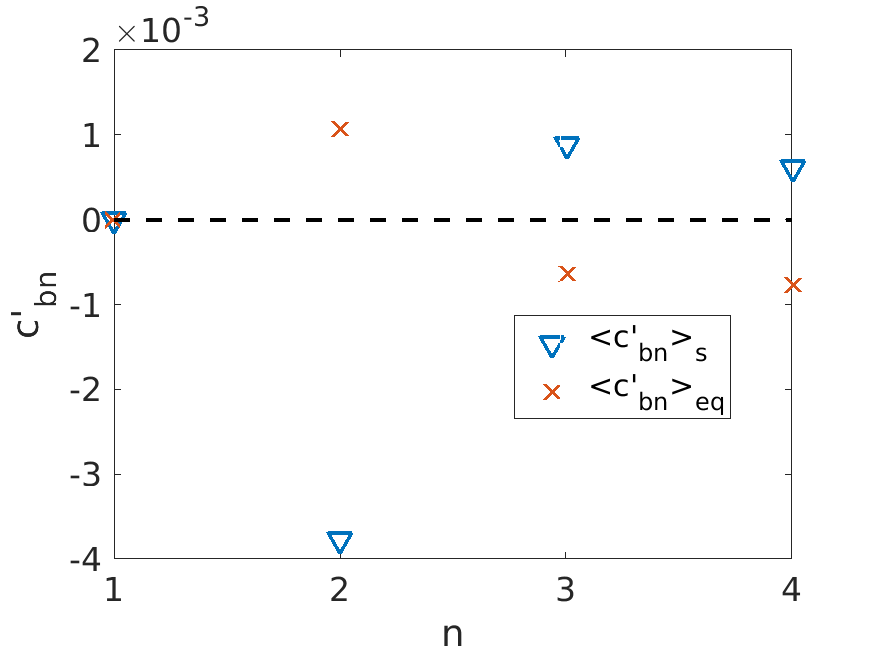} \\  
\end{tabular}
\caption{Left ) Relative contribution of the POD modes to the variations of the base  pressure coefficient (see text). Right) Contribution of the modes corresponding to the characteristic states $a_n^{p,eq}$ and $a_n^{p,s}$. }
\label{contributionpressure}
\end{figure}

%
%

\begin{figure}[h]
\hspace{-2in}

\begin{tabular}{llllll}
\includegraphics[ height=0.8cm]{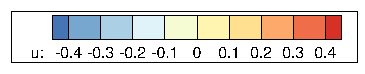}  & & 
\includegraphics[ height=0.8cm]{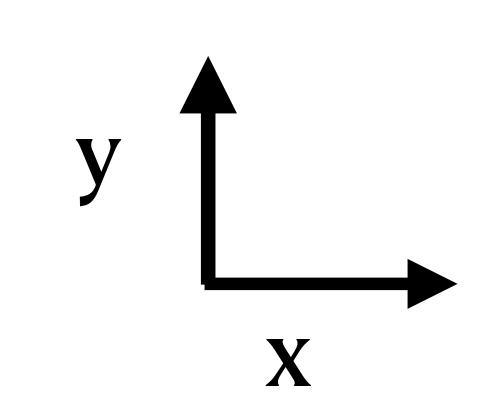}   
& \\
\includegraphics[trim=1.75cm  6cm 2.85cm 7cm, clip, height=1.9cm]{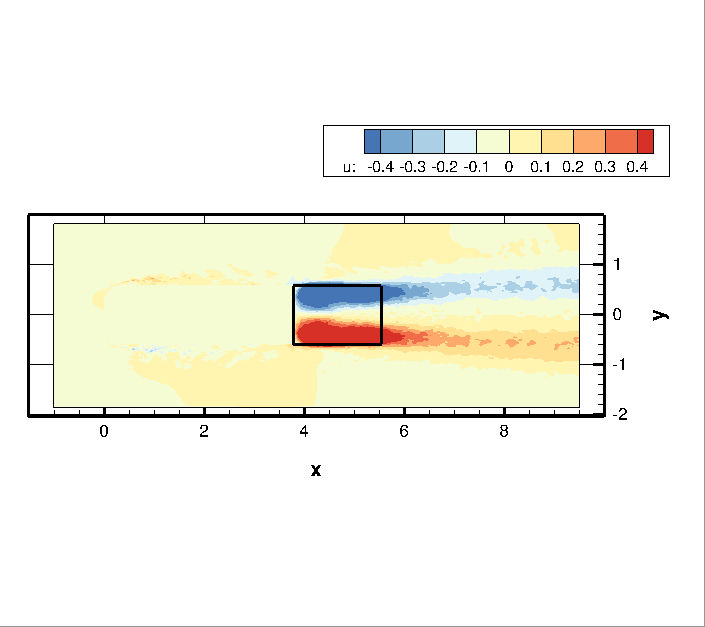} &
\includegraphics[trim=0cm  0cm 0cm 0cm, clip, height=1.9cm]{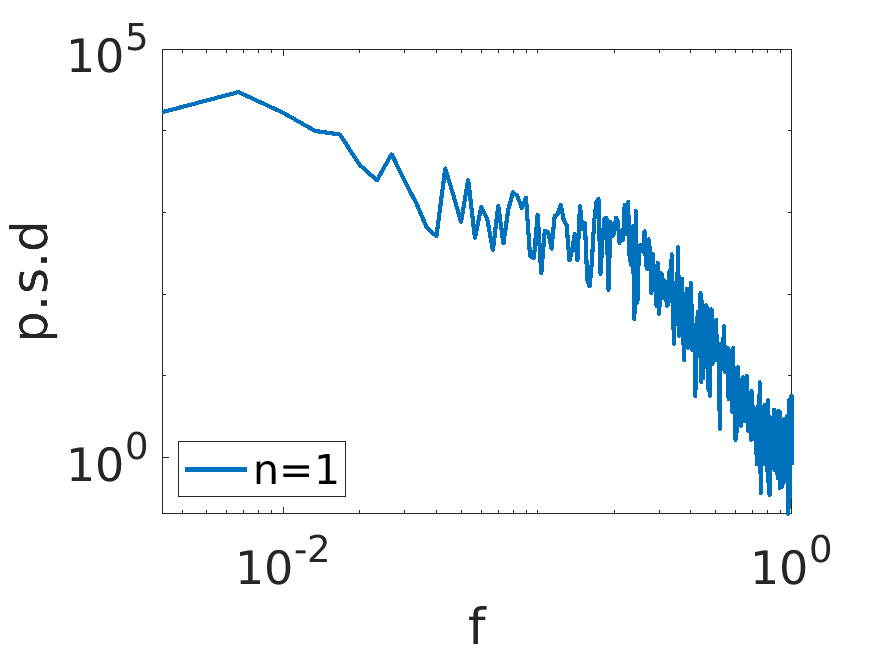} &
\includegraphics[trim=1.75cm  6cm 2.85cm 7cm, clip, height=1.9cm]{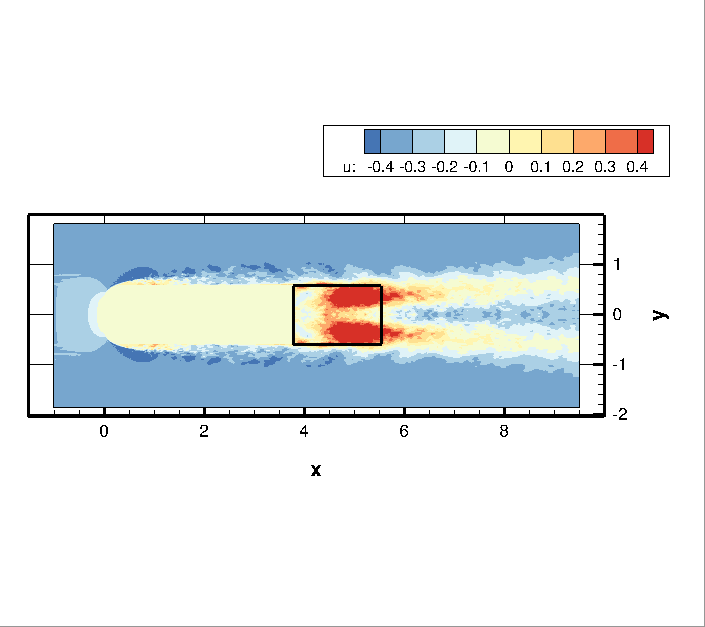} & 
\includegraphics[trim=0cm  0cm 0cm 0cm, clip, height=1.9cm]{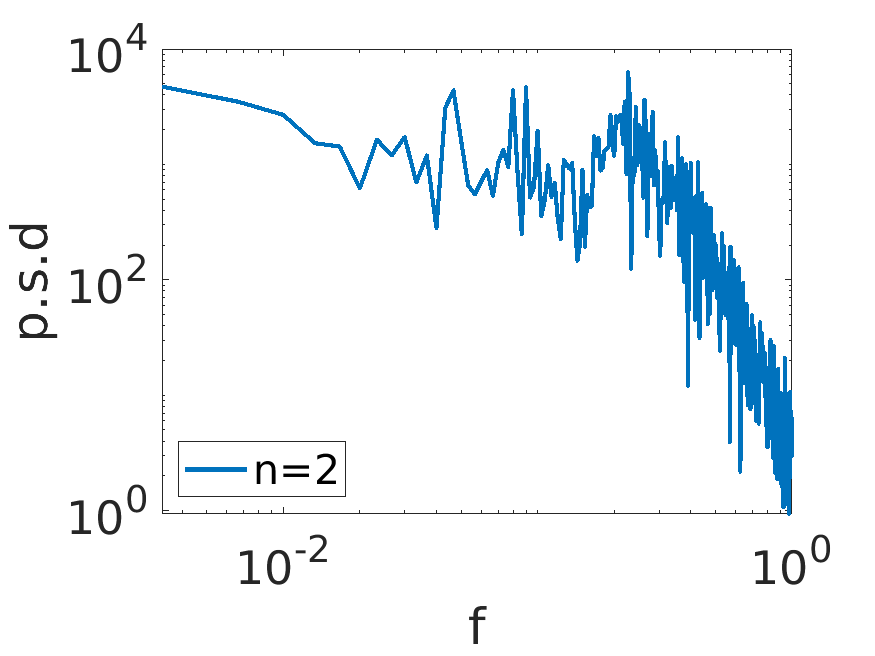} \\
\includegraphics[trim=1.75cm  6cm 2.85cm 7cm, clip, height=1.9cm]{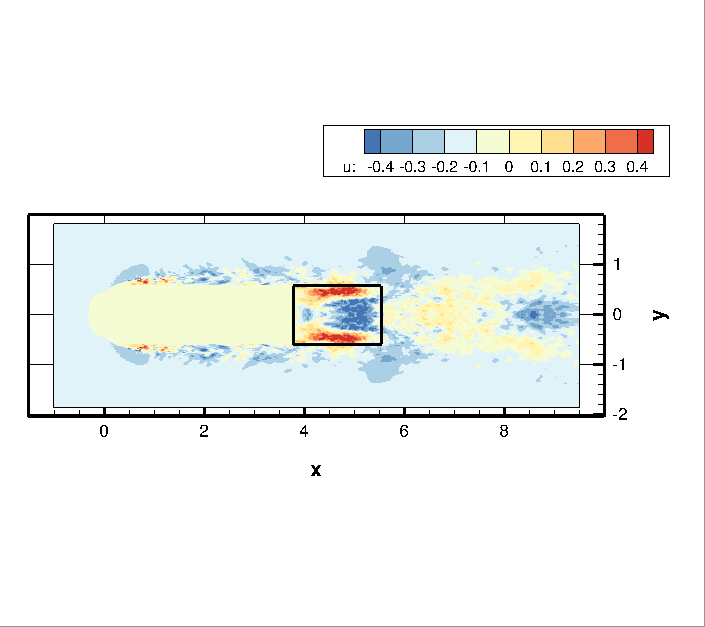} &
\includegraphics[trim=0cm  0cm 0cm 0cm, clip, height=1.9cm]{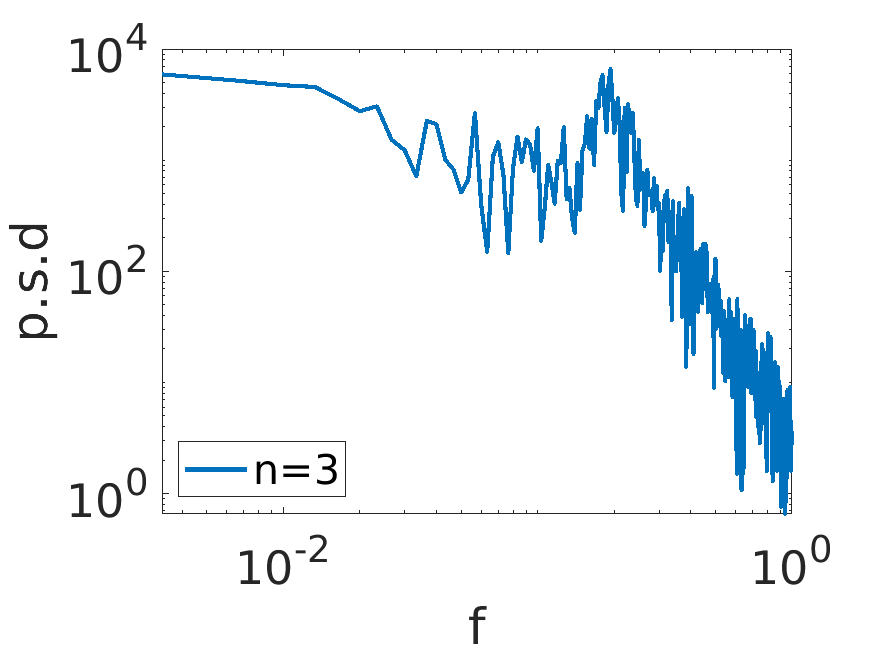} &
\includegraphics[trim=1.75cm  6cm 2.85cm 7cm, clip, height=1.9cm]{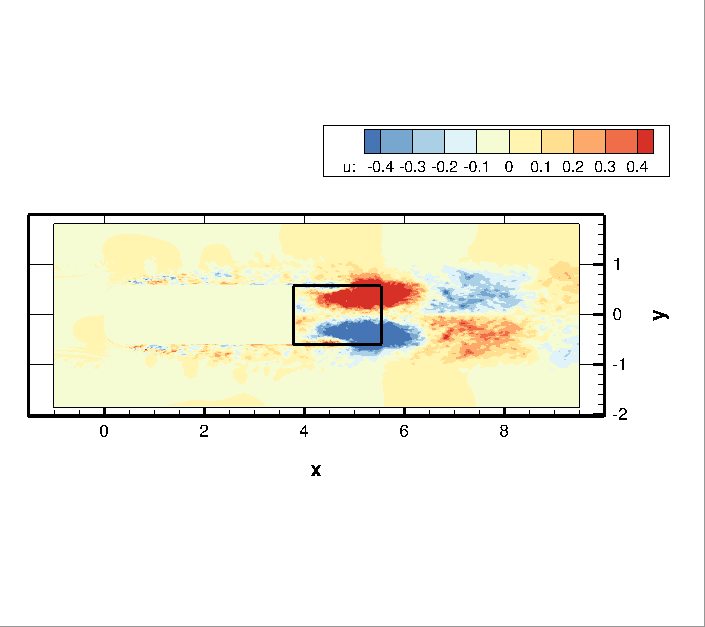} & 
\includegraphics[trim=0cm  0cm 0cm 0cm, clip, height=1.9cm]{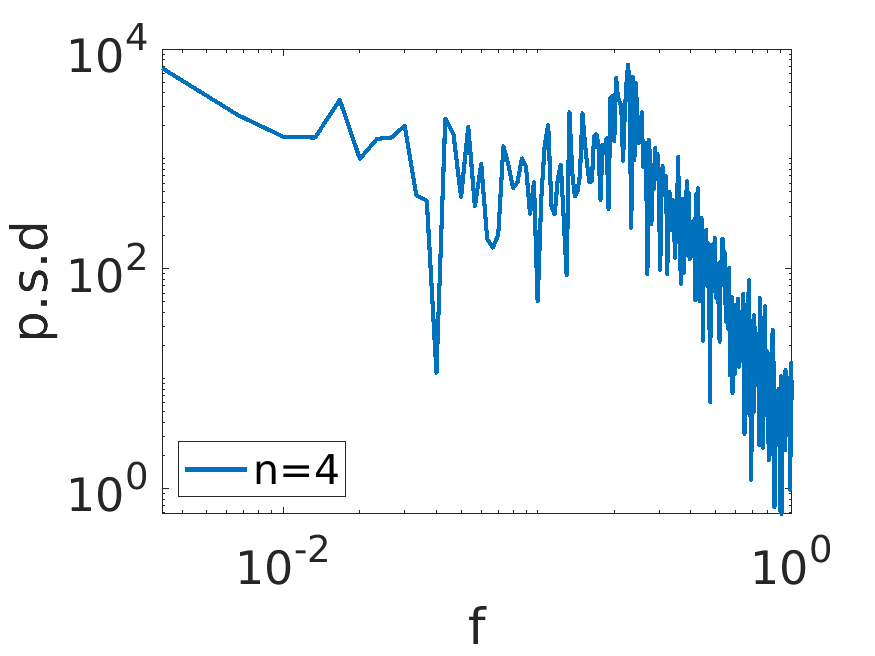} \\ 
\includegraphics[trim=1.75cm  6cm 2.85cm 7cm, clip, height=1.9cm]{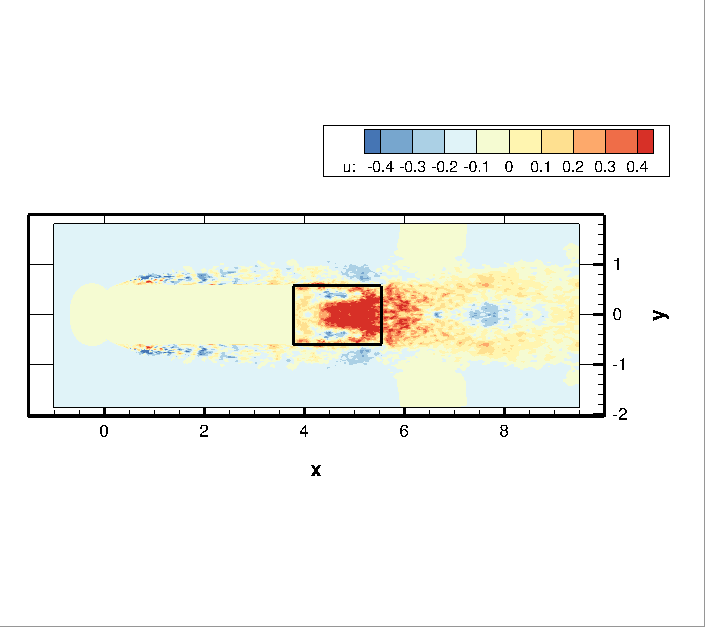} &
\includegraphics[trim=0cm  0cm 0cm 0cm, clip, height=1.9cm]{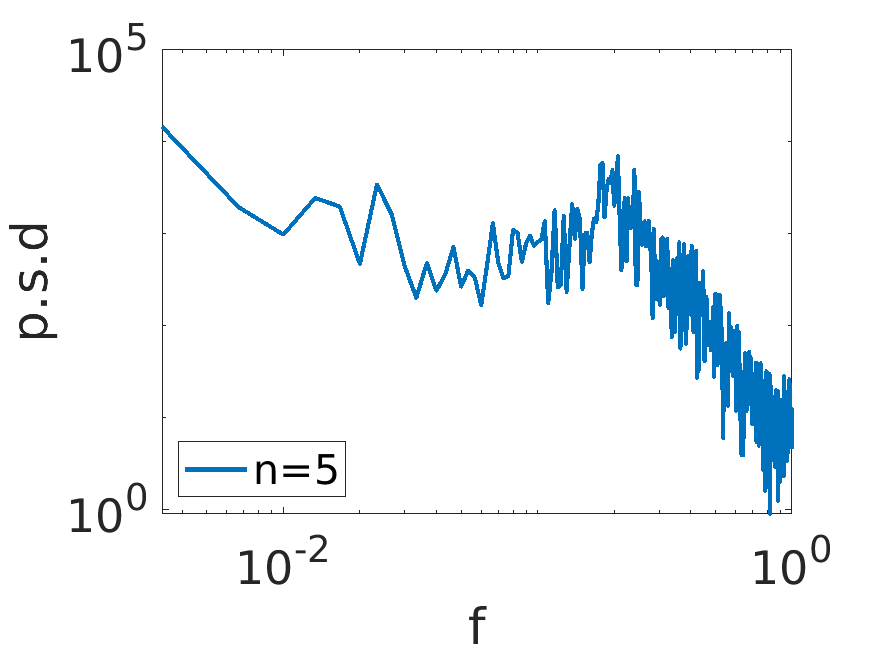} &
\includegraphics[trim=1.75cm  6cm 2.85cm 7cm, clip, height=1.9cm]{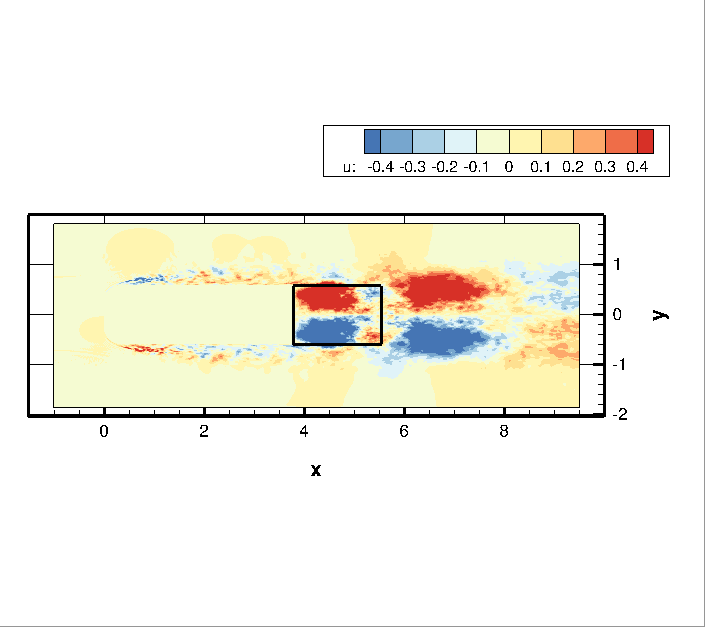} & 
\includegraphics[trim=0cm  0cm 0cm 0cm, clip, height=1.9cm]{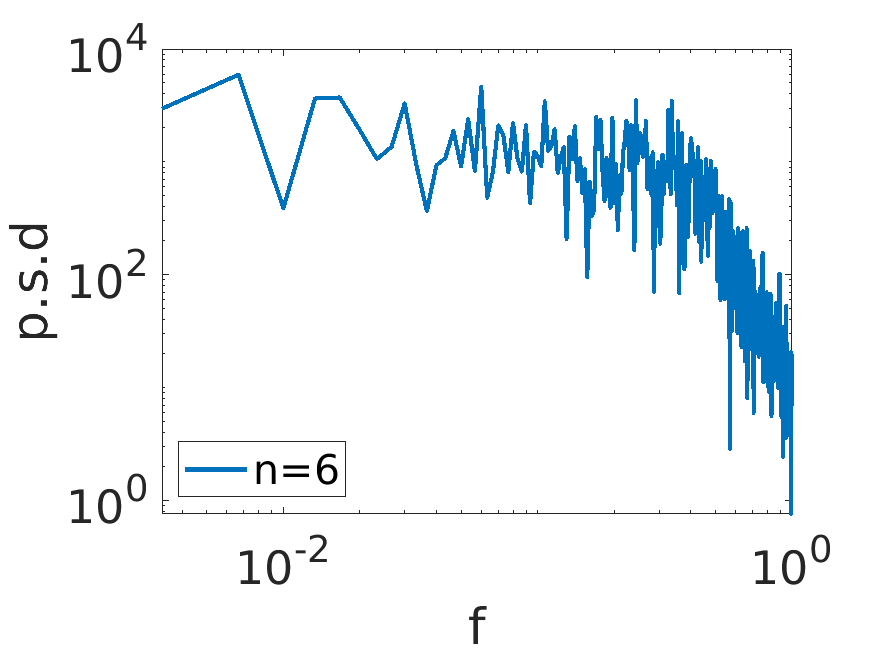} \\ 
\includegraphics[trim=1.75cm  6cm 2.85cm 7cm, clip, height=1.9cm]{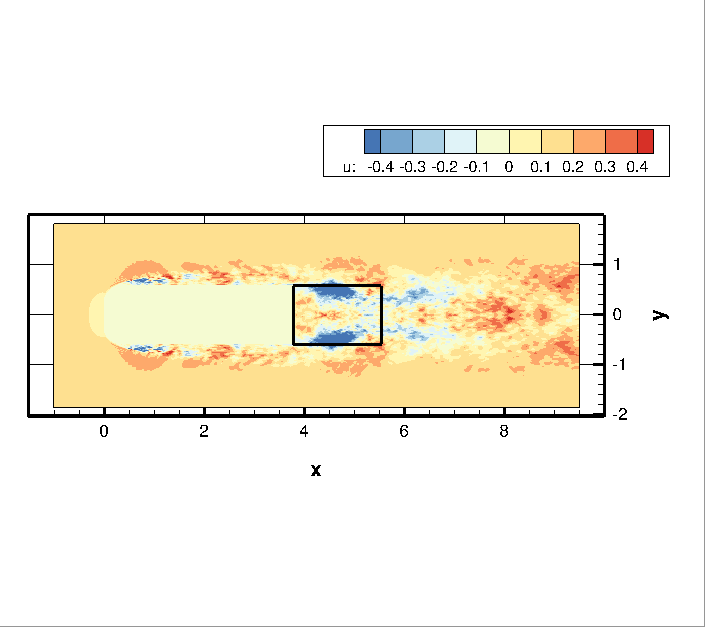} &
\includegraphics[trim=0cm  0cm 0cm 0cm, clip, height=1.9cm]{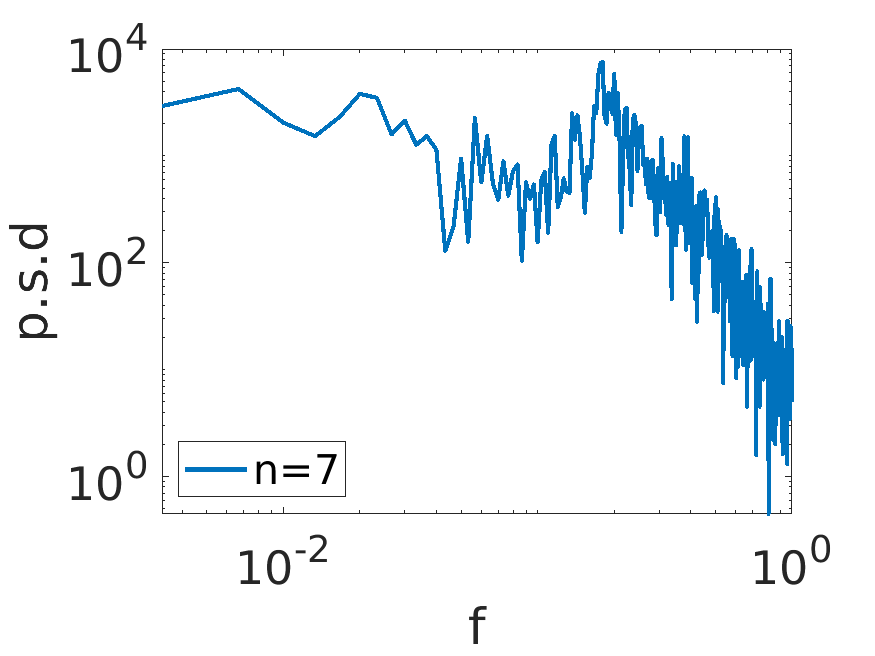} & 
\includegraphics[trim=1.75cm  6cm 2.85cm 7cm, clip, height=1.9cm]{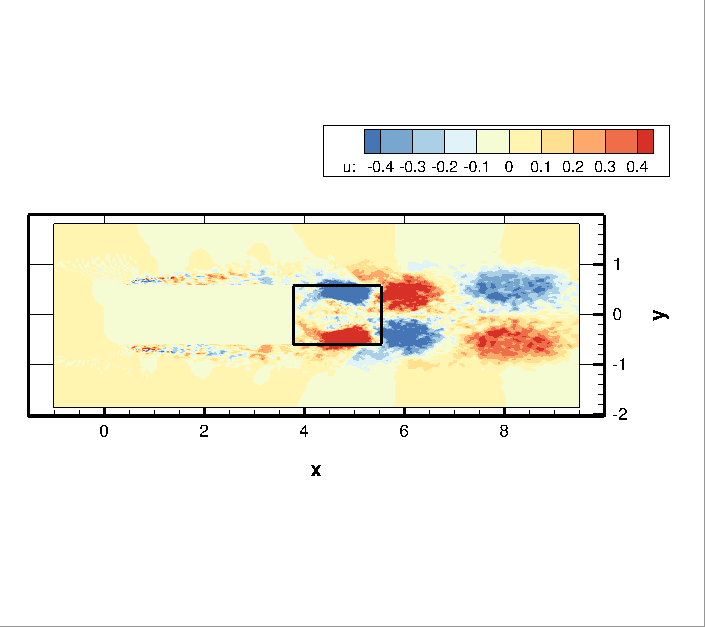} &  
\includegraphics[trim=0cm  0cm 0cm 0cm, clip, height=1.9cm]{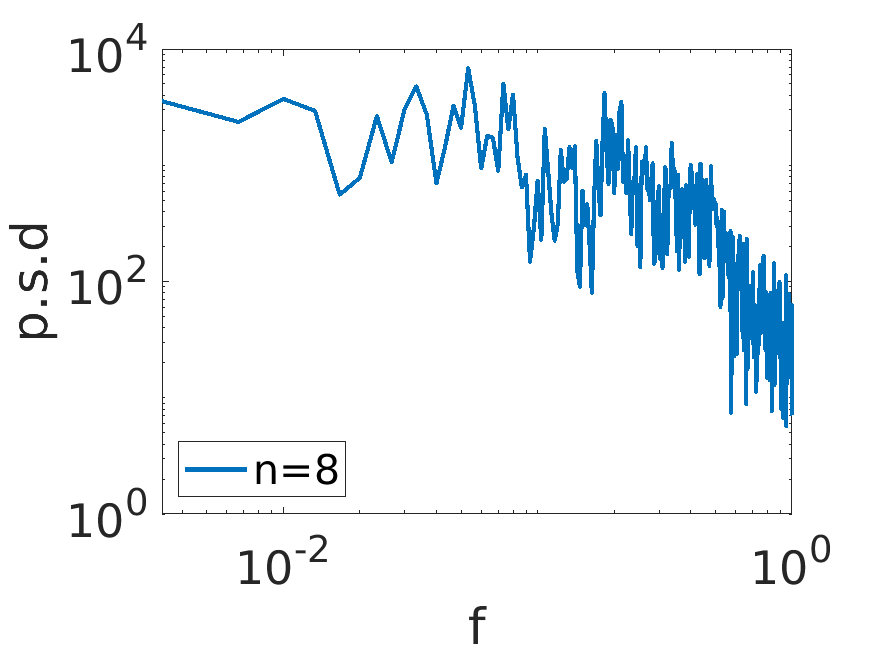}  \\
\includegraphics[trim=1.75cm  6cm 2.85cm 7cm, clip, height=1.9cm]{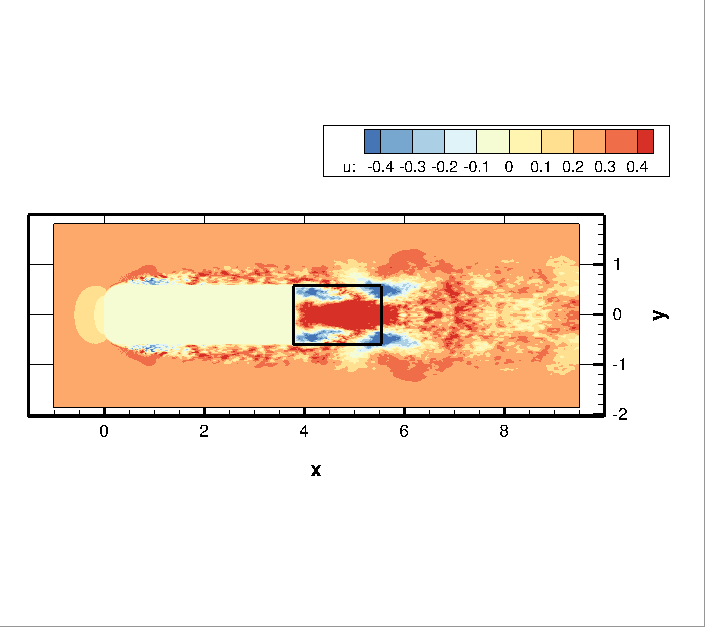}  &
\includegraphics[trim=0cm  0cm 0cm 0cm, clip, height=1.9cm]{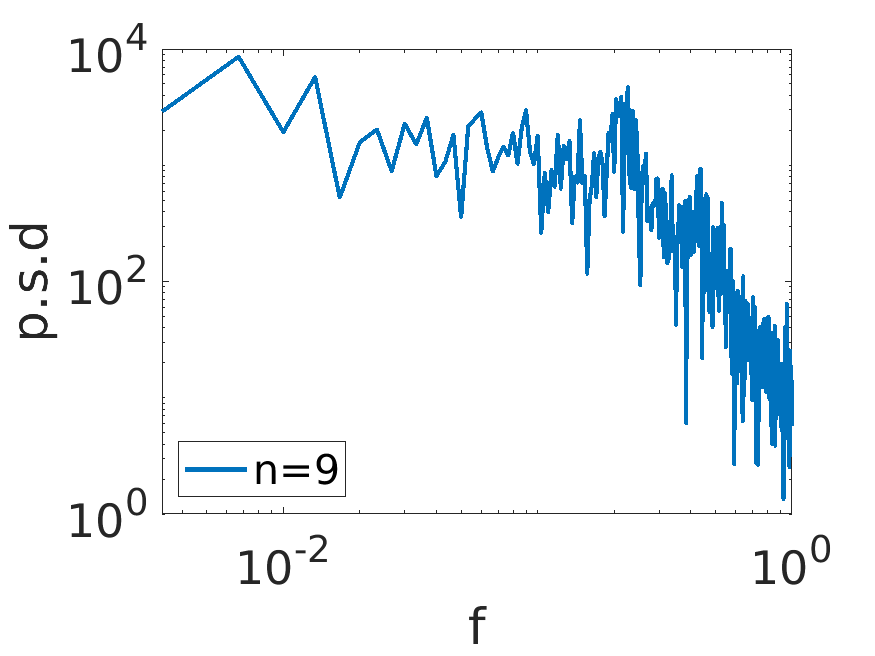} & 
\includegraphics[trim=1.75cm  6cm 2.85cm 7cm, clip, height=1.9cm]{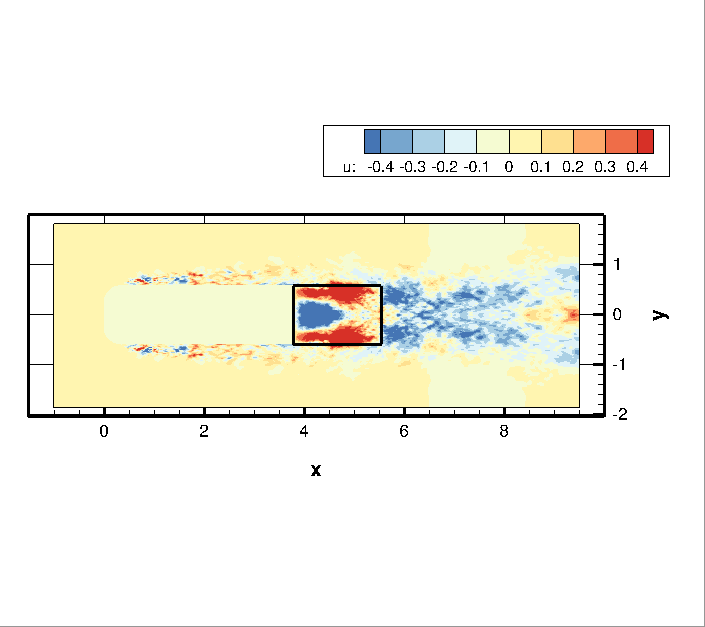}  & 
\includegraphics[trim=0cm  0cm 0cm 0cm, clip, height=1.9cm]{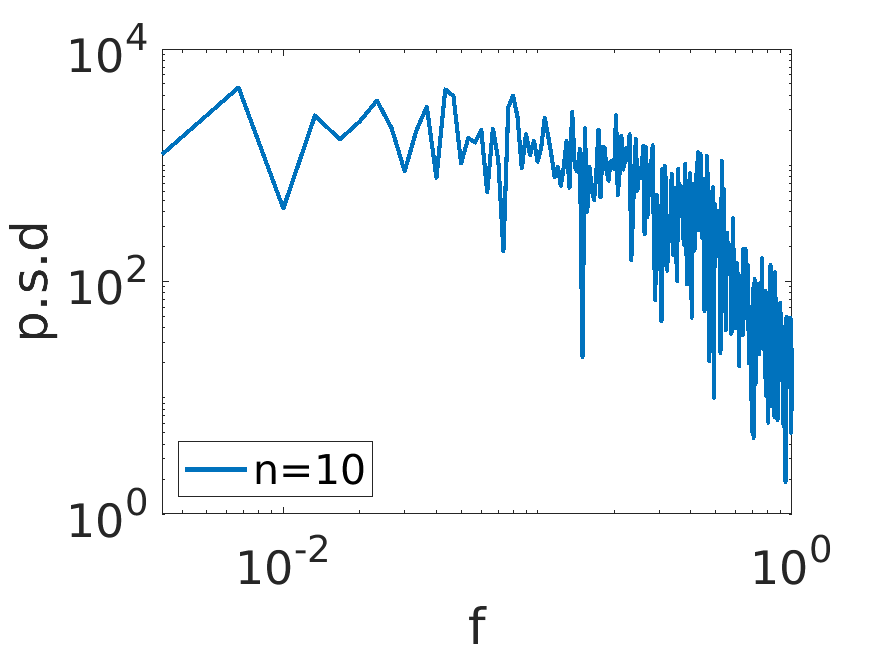}  \\
\end{tabular}

\caption{Velocity modes  computed over the near-wake  volume (delimited with the thick black line) and extended to the full domain, viewed in a horizontal plane at mid-height.}
\label{velmodes}
\end{figure}

\begin{figure}[h]
\hspace{-2in}
\begin{tabular}{ll}
\hspace{1.8in} a) &\hspace{1.8in}  b) \\
 \includegraphics[trim=0.cm 0cm 0.8cm 0cm, clip, height=5cm]{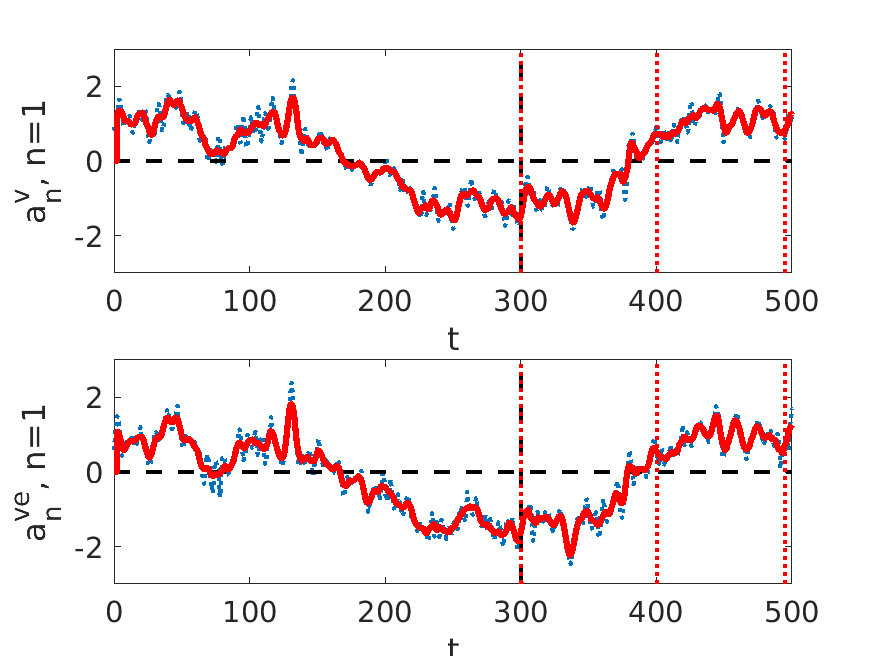} &  
 \includegraphics[trim=0.cm 0cm 0.8cm 0cm, clip, height=5cm]{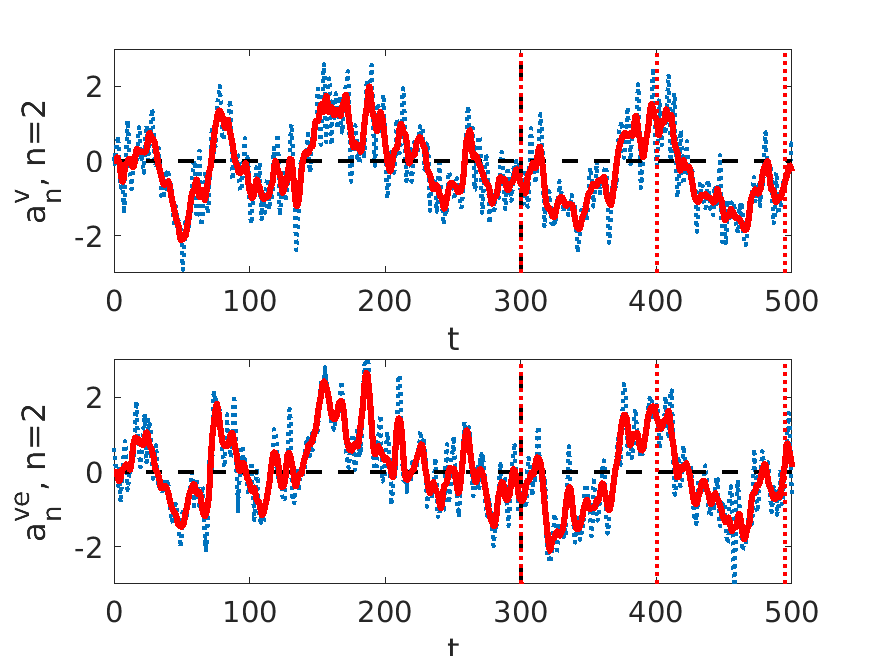} \\  
\hspace{1.8in} c) &\hspace{1.8in} d) \\
 \includegraphics[trim=0.cm 0cm 0.8cm 0cm, clip, height=5cm]{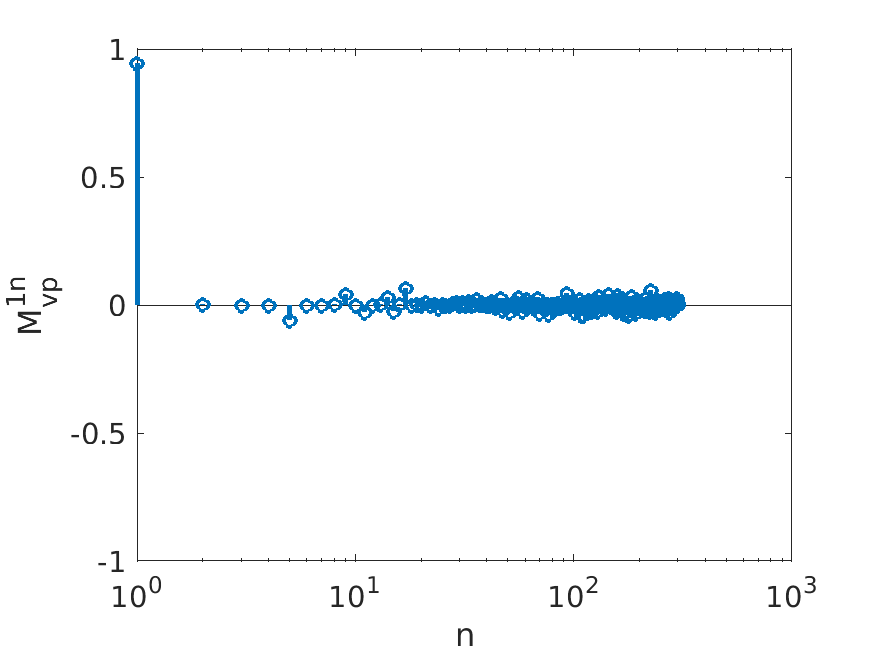}  & 
 \includegraphics[trim=0.cm 0cm 0.8cm 0cm, clip,height=5cm]{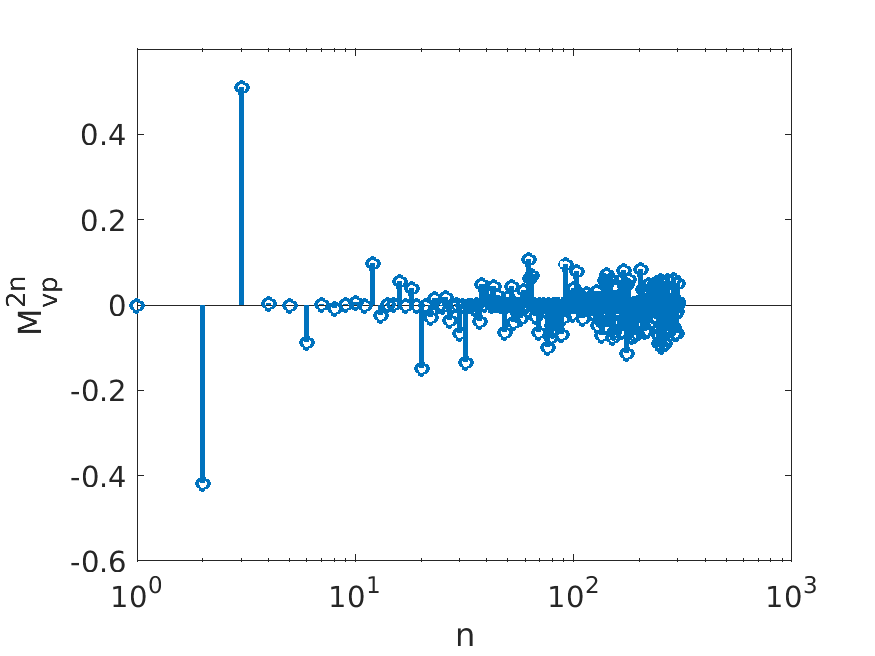} \\ 
\end{tabular}
\caption{
a)b): Velocity mode contribution to first POD mode $M_{1n}^{vp}$ (a) and second POD mode $M_{2n}^{vp}$ (b). 
c)d): Velocity POD amplitudes for the first (c) and 
second mode (d): the top row corresponds
to the exact coefficient $a_n^v$. 
The bottom row corresponds to the estimated coefficient $a_n^{v,e}$. 
The dotted line represents the instantaneous signal, and the red 
thick line to the signal filtered with a moving average of 5 time units.
The vertical dashed line corresponds to the limit of the POD snapshot acquisition.
The three red dotted lines correspond to the times selected for reconstruction 
in figure \ref{velrecons}.
 }
\label{estimvel12}
\end{figure}

\begin{figure}[h]
\hspace{-2in}
\begin{tabular}{ll}
\hspace{1.8in} a) &\hspace{1.8in}  b) \\
 \includegraphics[trim=0.cm 0cm 0.8cm 0cm, clip,height=5cm]{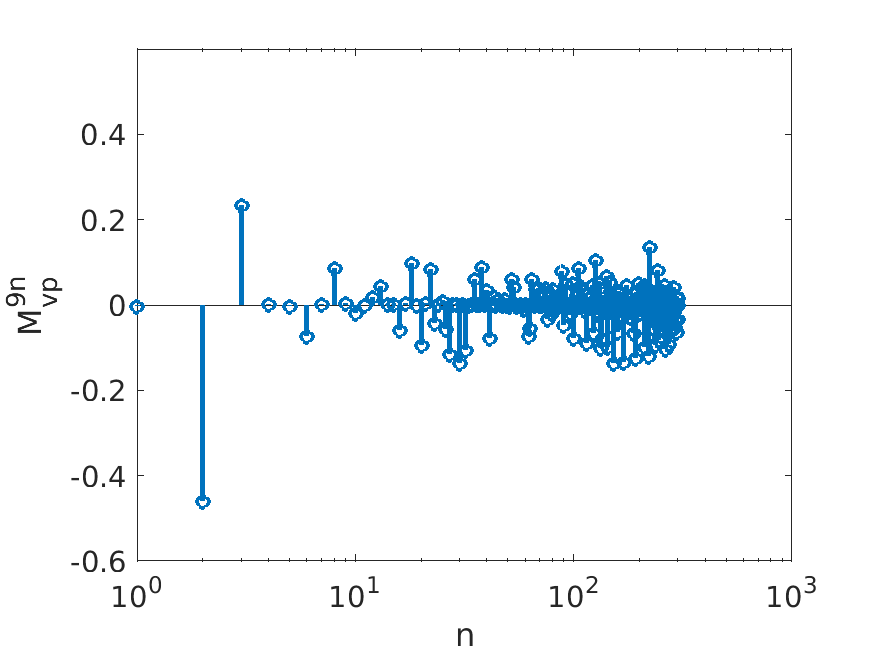}  & 
 \includegraphics[trim=0.cm 0cm 0.8cm 0cm, clip,height=5cm]{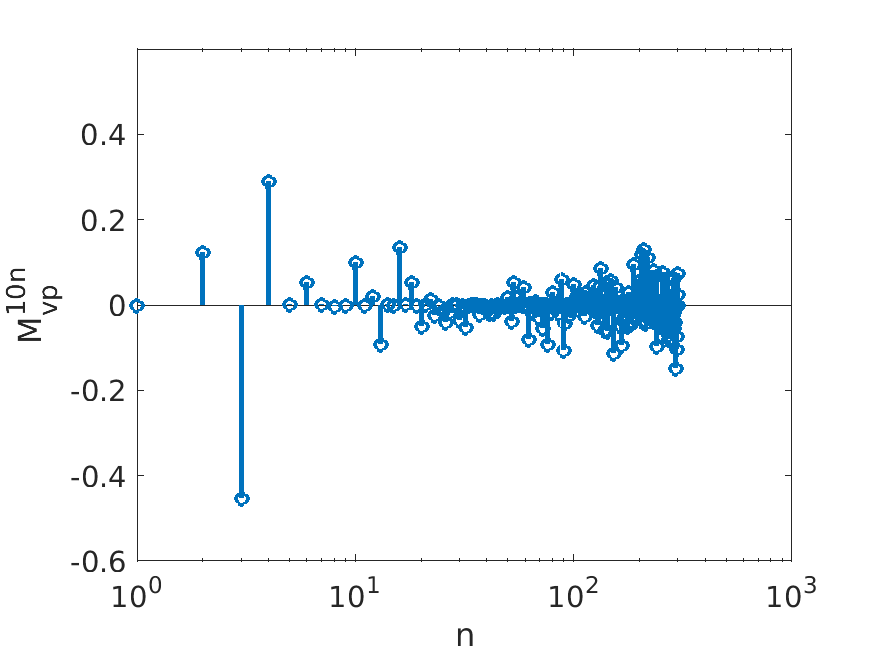} \\ 
\hspace{1.8in} c) &\hspace{1.8in}  d) \\
 \includegraphics[trim=0.cm 0cm 0.8cm 0cm, clip,height=5cm]{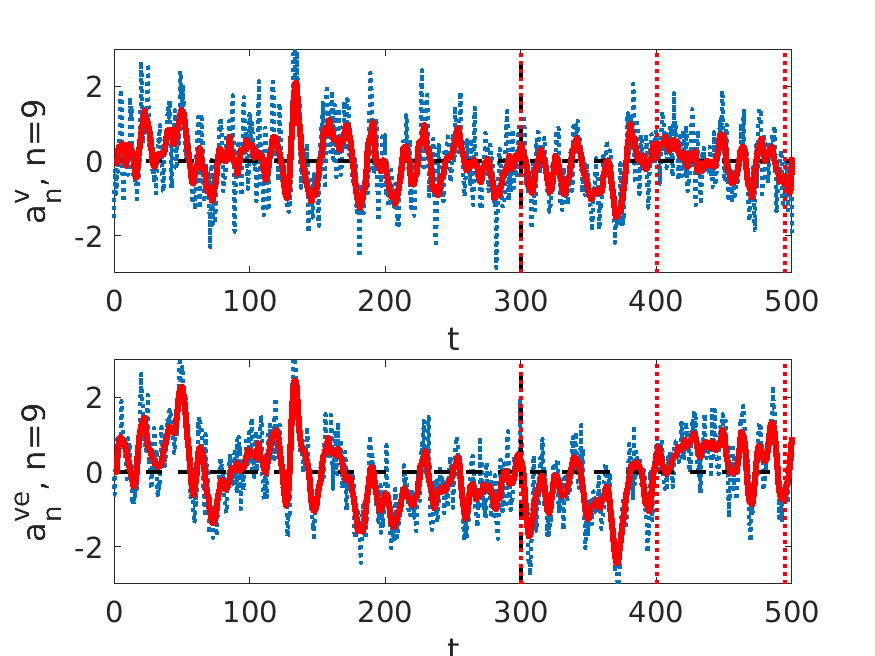}  & 
 \includegraphics[trim=0.cm 0cm 0.8cm 0cm, clip,height=5cm]{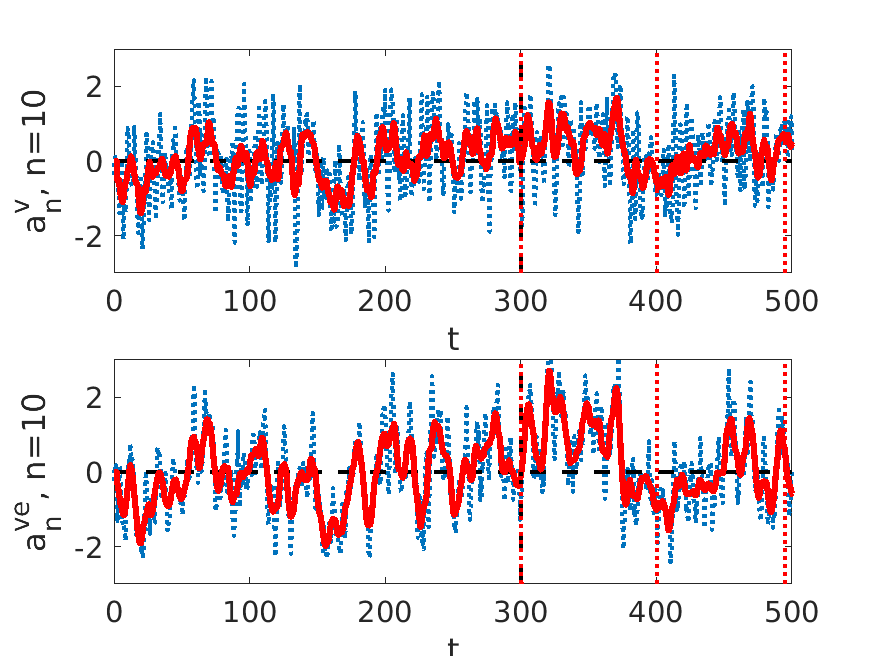} \\ 
\end{tabular}
\caption{
a)b): Influence coefficients  $M_{9n}^{vp}$ (a) and $M_{10n}^{vp}$ (b). 
c)d): Velocity POD amplitudes for the ninth (c) and 
tenth mode (d): legend as in figure \ref{estimvel12}}. 
\label{estimvel910}
\end{figure}

\begin{figure}[h]
\hspace{-1.8in}
\begin{tabular}{cccc}
\hspace{1.5in} $t=300$ & \hspace{1.5in} $t=385$ & \hspace{1.5in} $t=495$ & \\
\includegraphics[trim=0.5cm 0 3.5cm 0, clip, height=4.cm]{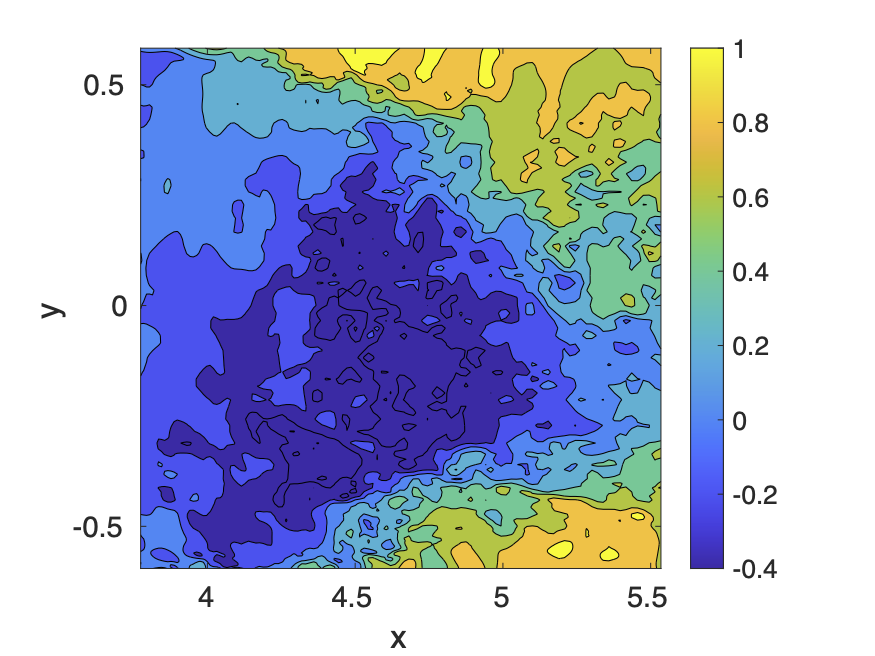}  & 
\includegraphics[trim=0.5cm 0 3.5cm 0, clip, height=4.cm]{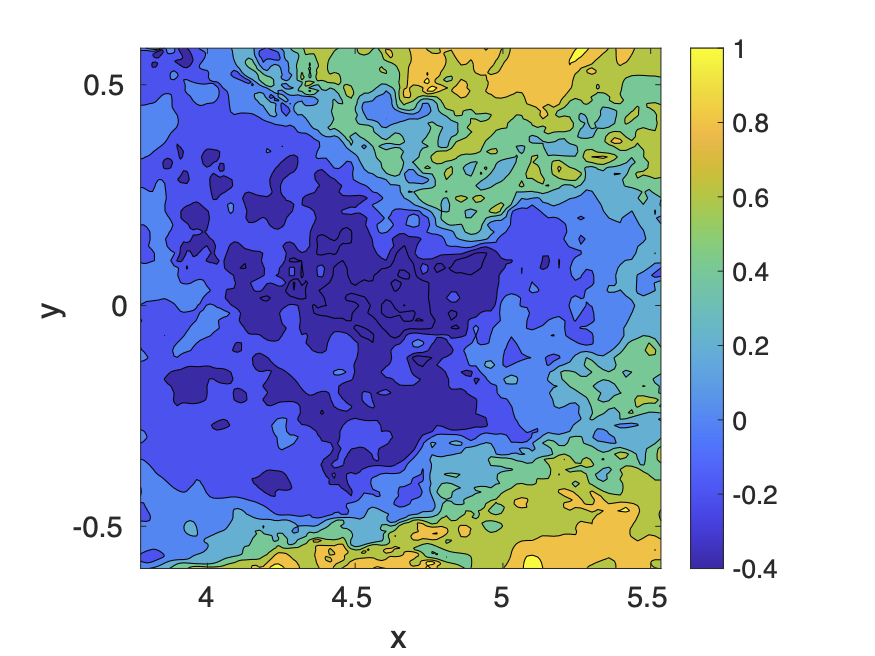}  & 
\includegraphics[trim=0.5cm 0 3.5cm 0, clip, height=4.cm]{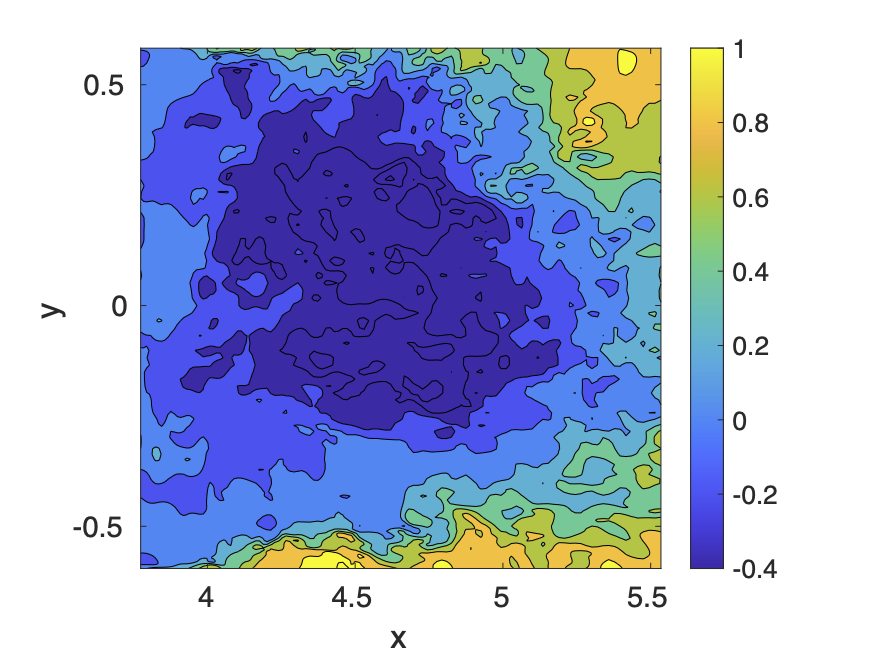} &
\hspace{1in} \includegraphics[trim=0.cm 0 0.cm 0, clip, height=4.cm]{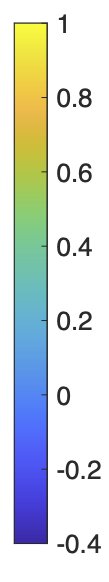} 
 \\ 
\includegraphics[trim=0.5cm 0 3.5cm 0, clip, height=4.cm]{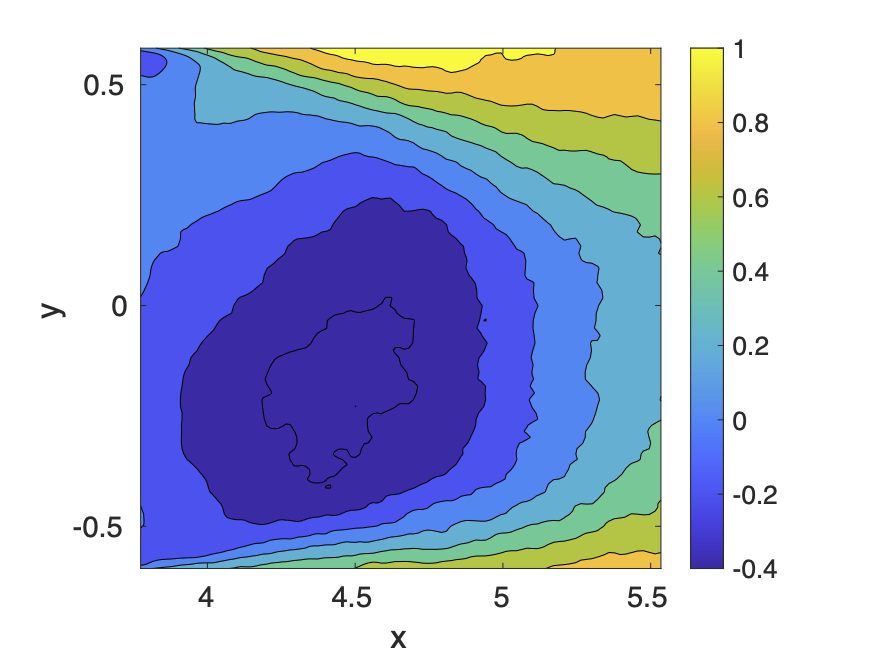}  & 
\includegraphics[trim=0.5cm 0 3.5cm 0, clip, height=4.cm]{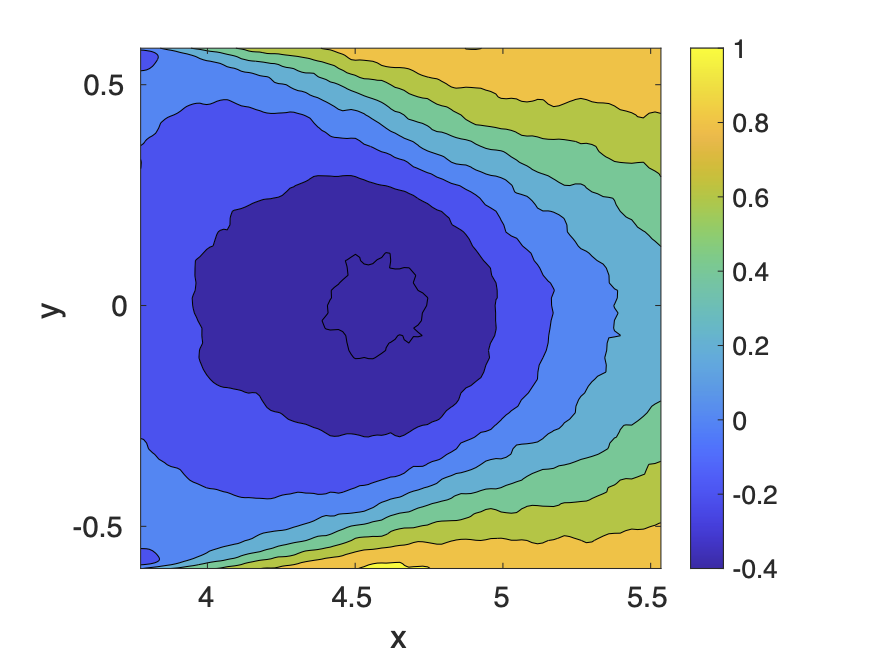}  & 
\includegraphics[trim=0.5cm 0 3.5cm 0, clip, height=4.cm]{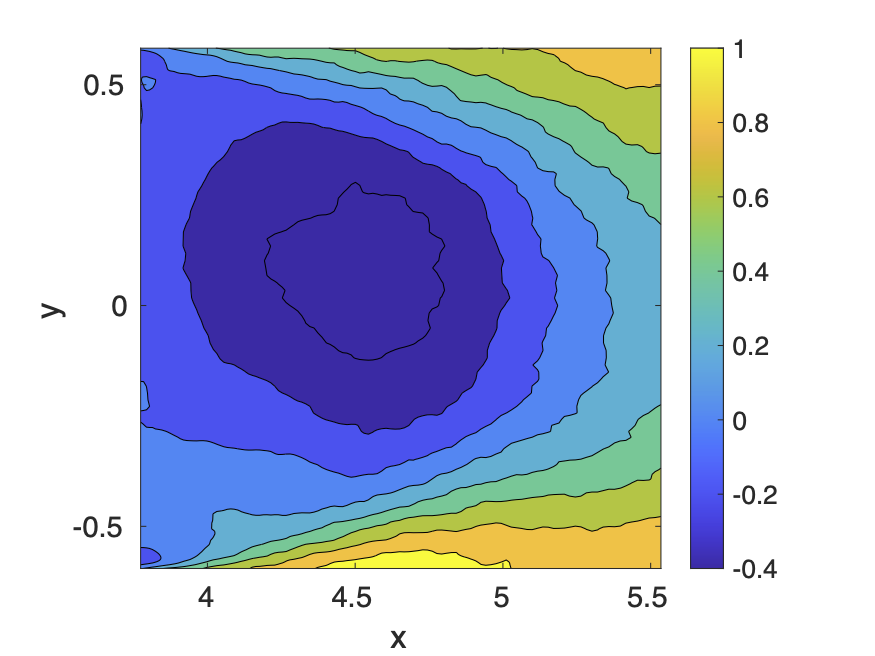}  & \\ 
\includegraphics[trim=0.5cm 0 3.5cm 0, clip, height=4.cm]{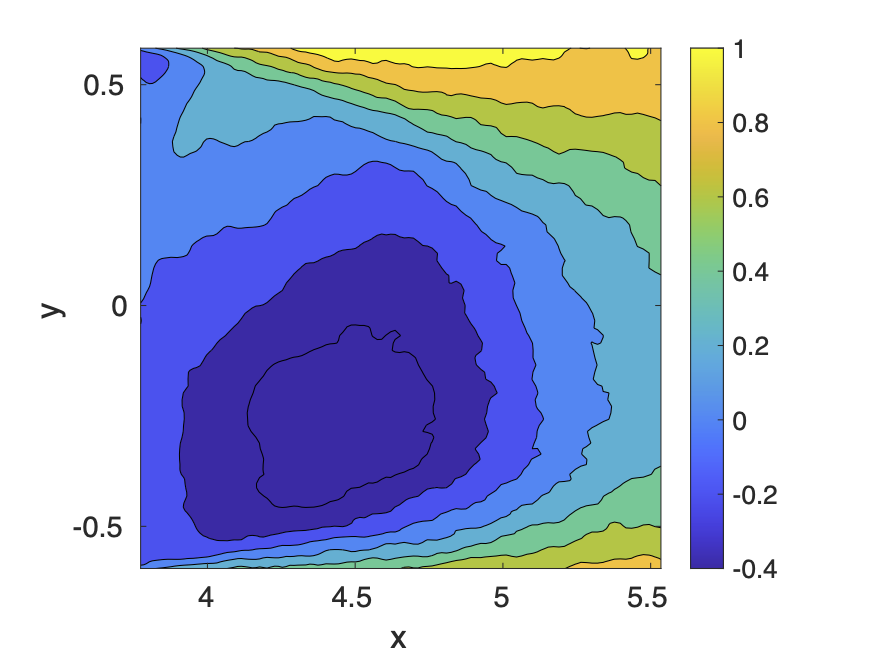}  & 
\includegraphics[trim=0.5cm 0 3.5cm 0, clip, height=4.cm]{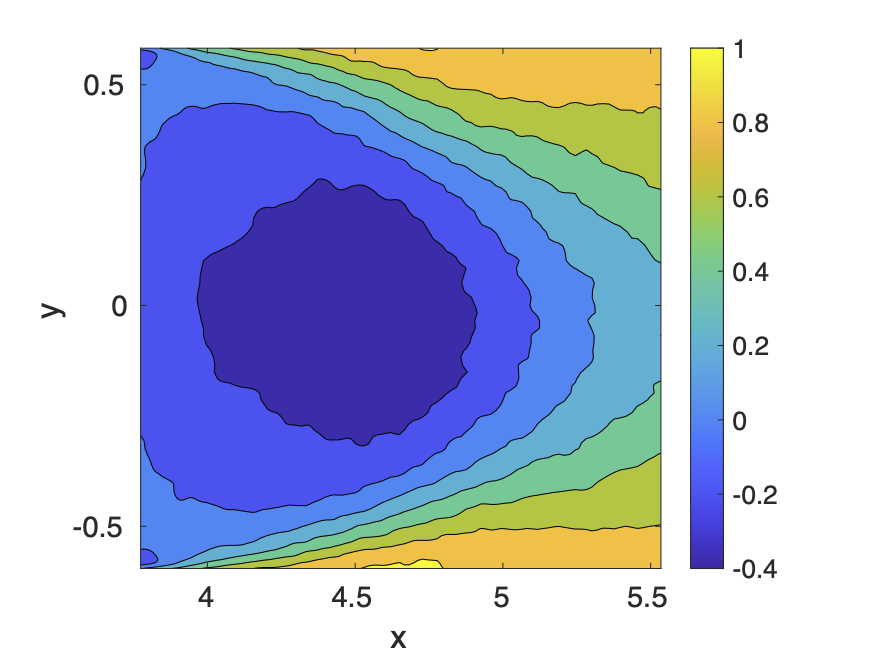}  & 
\includegraphics[trim=0.5cm 0 3.5cm 0, clip, height=4.cm]{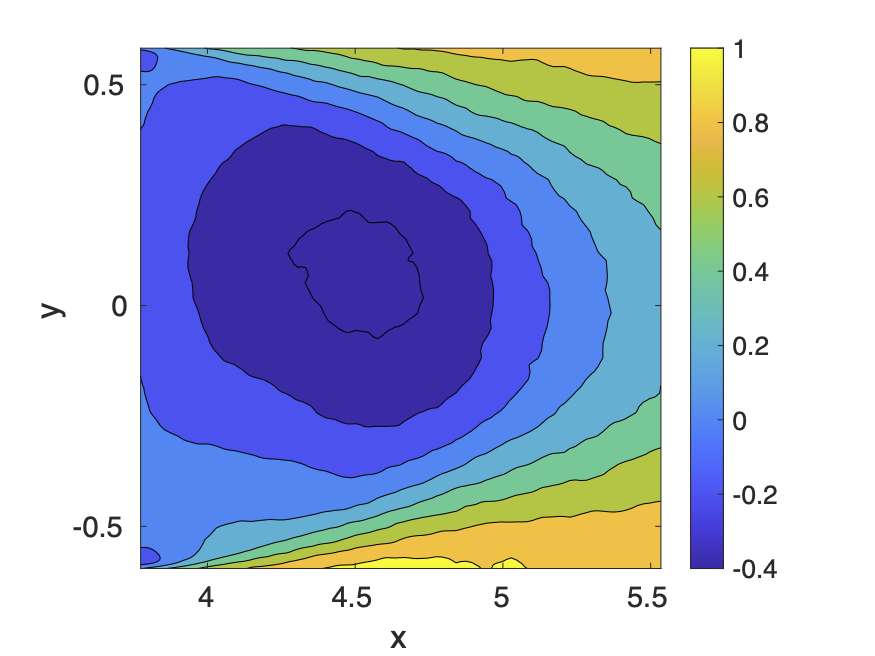}  & \\ 
\end{tabular}
\caption{ Streamwise velocity field at three instants $t=  300, 385, 495$ materialized by the vertical dotted lines in figures \ref{estimvel12} and \ref{estimvel910} ;
Top row: DNS field; Middle row: Field projected on the first 10 POD velocity modes; Bottom row: Field estimated from the first 10 POD pressure modes.}
\label{velrecons}
\end{figure}

\begin{table}
\begin{tabular}{|l|l|l|l|l|l|l|l|l|l|l|} \hline
$C(a_n^v, a_n^{v e}) $ & $n=1$ & $n=2$ & $n=3$ & $n=4$ & $n=5$ & $n=6$ & $n=7$ & $n=8$ & $n=9$ & $n=10$ \\ \hline
instantaneous  &   0.94 & 0.67 & 0.30 & 0.16 & 0.04 & 0.31 & 0.30 & 0.35 & 0.40 & 0.57 \\  \hline
averaged over $\tau=5$ & 0.99 & 0.93 & 0.27 & 0.62 & 0.08 & 0.52 & 0.41 & 0.60 & 0.73 & 0.81 \\   \hline
\end{tabular}
\caption{Correlation coefficient between the velocity POD amplitude and the pressure-based estimation ; top row: for instantaneous amplitudes; bottom row:
for amplitudes filtered with a moving average of 5 convective time units.}
\label{coefcorrelvel}  
\end{table}

\begin{table}
\begin{tabular}{|l|l|l|l|l|l|} \hline
$C(a_n^p, a_n^{p e})$ & $n=1$ & $n=2$ & $n=3$ & $n=4$ & $n=5$  \\ \hline
instantaneous  &   0.94 & 0.67 & 0.82 & 0.31 & 0.37  \\  \hline
averaged over $\tau=5$ & 0.99 & 0.86 & 0.91 & 0.44 & 0.80  \\   \hline
\end{tabular}
\caption{Correlation coefficient between the pressure POD amplitude and the velocity-based estimation; top row: for instantaneous amplitudes; bottom row:
for amplitudes filtered with a moving average of 5 convective time units.}
\label{coefcorrelpres} 
\end{table}

\begin{figure}[h]
\hspace{-2in}
\begin{tabular}{ll}
\hspace{1.8in} a) &\hspace{1.8in}  b) \\
\includegraphics[trim=0.cm 0cm 0.8cm 0cm, clip,height=5cm]{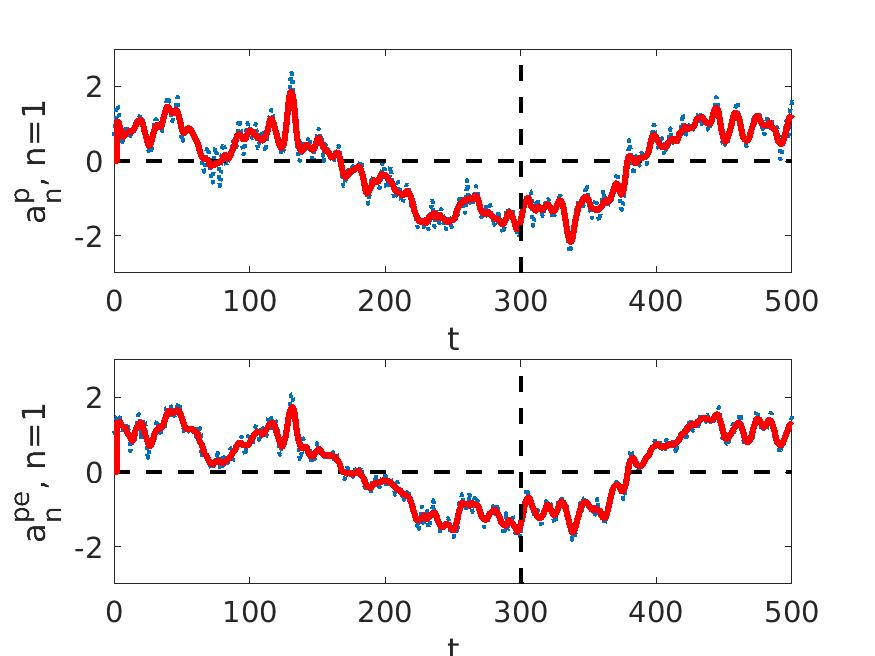}  & 
 \includegraphics[trim=0.cm 0cm 0.8cm 0cm, clip,height=5cm]{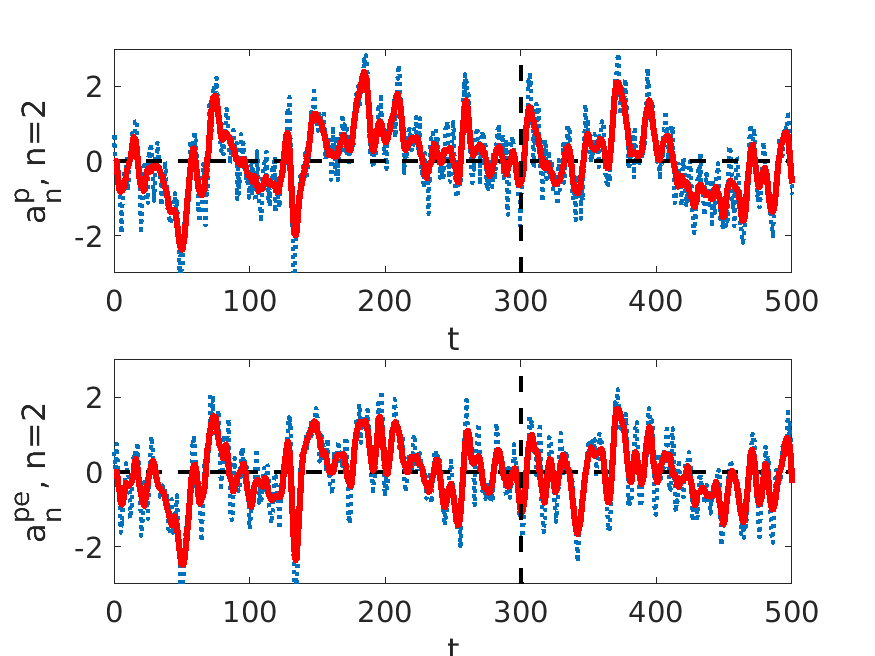} \\ 
\hspace{1.8in} c) &\hspace{1.8in}  d) \\
\includegraphics[trim=0.cm 0cm 0.8cm 0cm, clip,height=5cm]{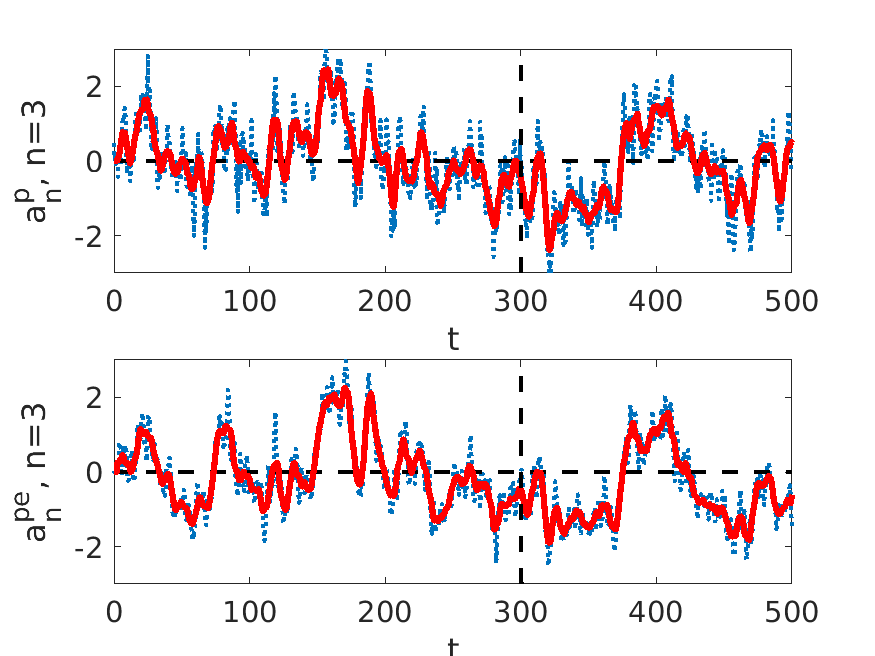}  & 
 \includegraphics[trim=0.cm 0cm 0.8cm 0cm, clip,height=5cm]{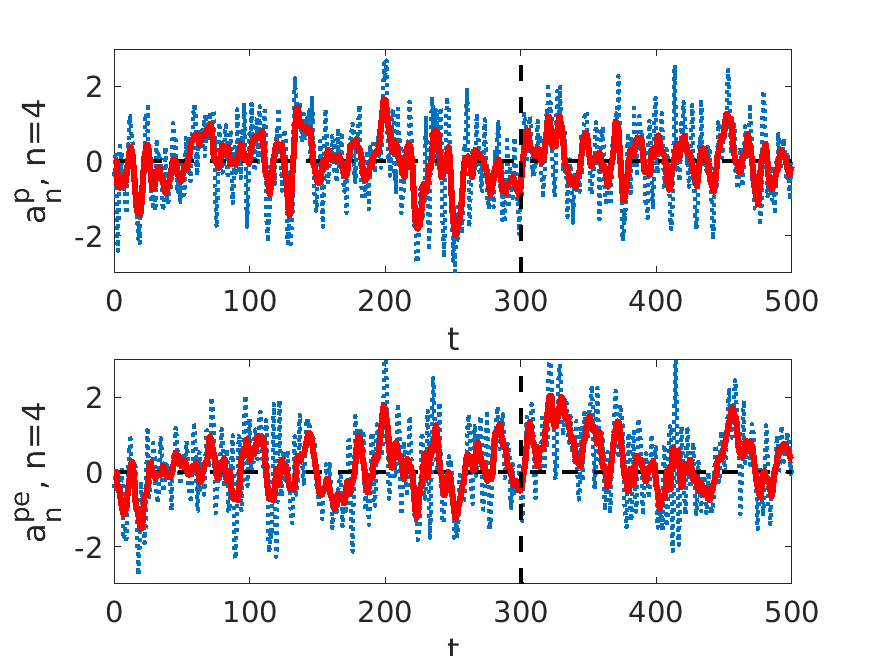} \\ 
\end{tabular}
\caption{Base pressure POD coefficients estimated from the near-wake velocity field.  The top row of each subfigure corresponds
to the exact coefficient $a_n^p$.
The bottom row corresponds to the estimated coefficient $a_n^{p,e}$.
The dotted line represents the instantaneous coefficient, and the red
thick line to its filtered with a moving average of 5 time units. The  dashed vertical line corresponds to the limit of the POD snapshot acquisition.
a) $n=1$ b) $n=2$ c) $n=3$ d) $n=4$. 
  }
\label{estimpres}
\end{figure}

\begin{figure}[h]
\begin{tabular}{cc}
\hspace{-2in}
\includegraphics[height=5cm]{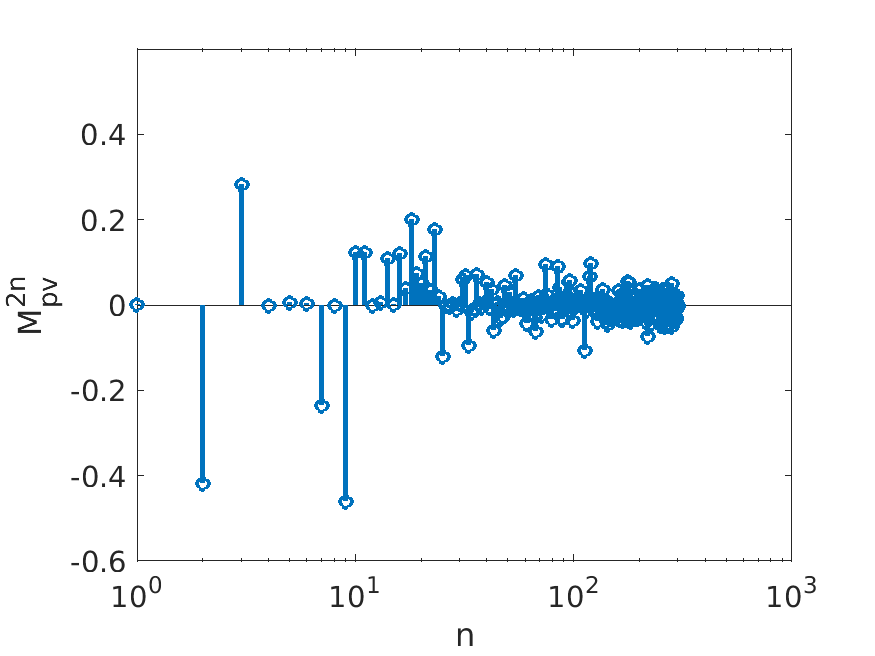}  & 
\includegraphics[height=5cm]{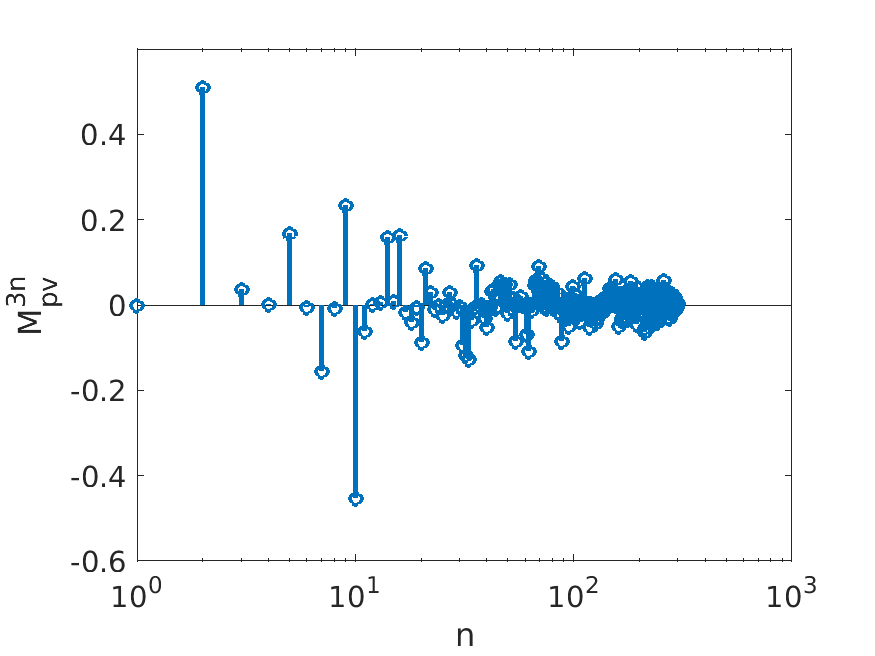} \\ 
\end{tabular}
\caption{Pressure contribution to second velocity mode  $M_{2n}^{pv}$ (a) and third velocity mode  $M_{3n}^{pv}$ (b) }
\label{Mpv23}
\end{figure}

\begin{figure}[h]
\hspace{-4in}
\includegraphics[height=8cm]{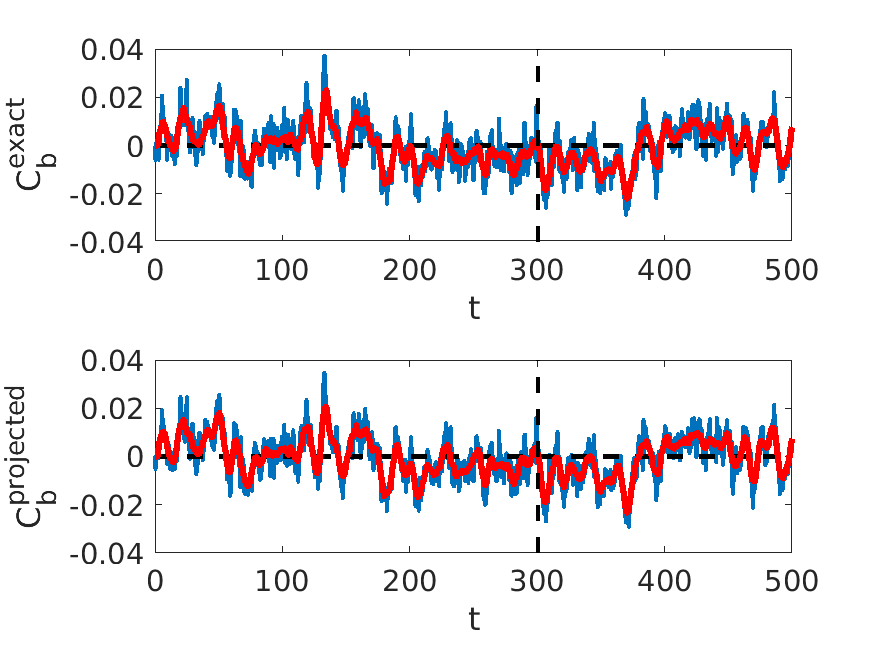} \\  
\hspace{-4in}
\includegraphics[height=8cm]{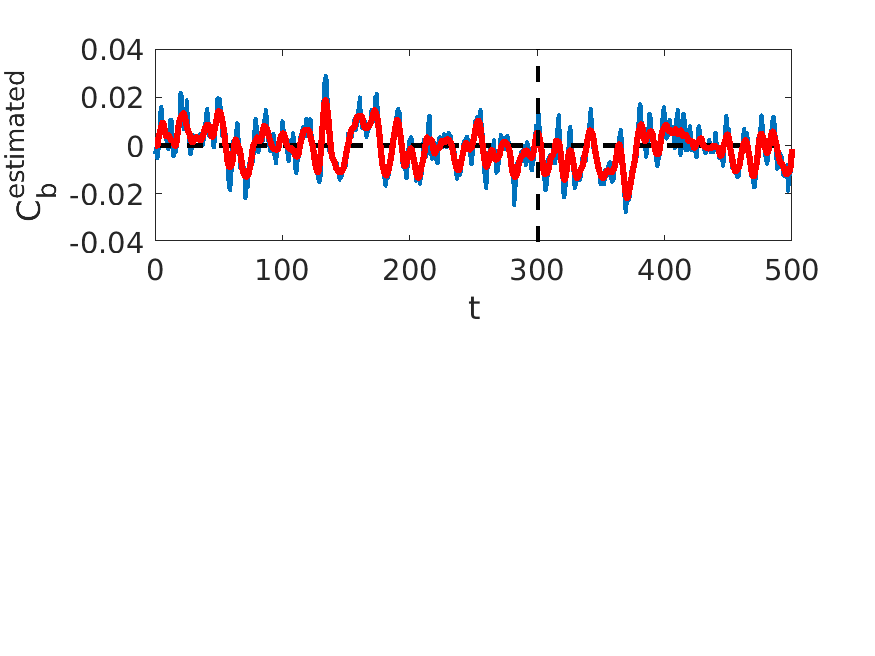} \\  
\caption{Base drag coefficient: top: DNS; middle: reconstruction 
based on pressure modes 2 and 3; bottom:  estimation based on 
POD velocity modes. The dashed vertical line 
corresponds to the limit of the POD snapshot acquisition.} 
\label{dragcoef}
\end{figure}

\end{document}